\documentclass[aps,twocolumn,preprintnumbers,raggedbottom,prb,amsmath,showpacs,nobalancelastpage]{revtex4}

\usepackage{epsfig,psfrag}
\usepackage{dcolumn}     
\usepackage{bm}          

\begin{document}

\title{Correlated sequential tunneling through a double barrier for interacting
one-dimensional electrons} 

\author{ M.\ Thorwart$^1$,  R.\ Egger$^1$, and M.\ Grifoni$^2$}

\affiliation{$^1$Institut f\"ur Theoretische Physik, 
Heinrich-Heine-Universit\"at D\"usseldorf,
D-40225 D\"usseldorf, Germany \\
$^2$ Institut f\"ur Theoretische Physik, Universit\"at Regensburg,
D-93040 Regensburg, Germany}

\date{\today}

\begin{abstract}
The problem of resonant tunneling through a quantum dot weakly coupled
to spinless Tomonaga-Luttinger liquids has been studied. We
compute the linear conductance due to sequential tunneling processes upon 
employing a master equation approach.  Besides the previously used 
lowest-order golden rule rates describing uncorrelated sequential 
tunneling (UST) processes, we systematically 
include higher-order correlated sequential tunneling
(CST) diagrams within the standard Weisskopf-Wigner approximation. 
We provide estimates for the parameter regions where CST effects can
be important.
Focusing mainly on the temperature dependence of the peak conductance, we
discuss the relation of these findings to previous theoretical and
experimental results.
\end{abstract}


\maketitle

\section{Introduction}

The startling properties of one-dimensional (1D) interacting electrons,
commonly referred to as (non-chiral) 
Tomonaga-Luttinger liquid (TLL) behavior,\cite{Tomonaga50,Luttinger63} see 
Refs.~\onlinecite{sh02,Giamarchi04} for reviews,   
have  recently moved into the focus of attention of the mesoscopic
 physics community.  
This was in particular prompted by 
 the successful demonstration of electrical transport experiments
for a variety of 1D materials, 
such as semiconductor quantum wires,\cite{auslaender} 
fractional quantum Hall edge states,\cite{chang} and
single-wall carbon nanotubes (SWNTs).\cite{dekker} 
In this paper,  we mainly focus on electrical transport in
SWNTs to keep the discussion
concrete.  Nevertheless, our results apply also to other systems.
Shortly after the theoretical proposal of TLL behavior in
individual metallic SWNTs,\cite{egger,kane}
the first experimental evidence for this
peculiar many-body state was 
reported.\cite{bockrath,yao,Nygard99,Postma00}
The expected TLL power-law scaling in the energy-dependent 
tunneling density of states
\cite{Kane92} in SWNTs has been verified experimentally.
In later experiments, \cite{Postma01,Bozovic01}
transport through an intrinsic
quantum dot formed by a double barrier within the SWNT
has been probed, allowing one to study the well-known
resonant or sequential tunneling  including Coulomb blockade
phenomena,\cite{Kulik75,Beenakker91,in92} but
 now for the case of leads formed by strongly correlated electrons.
When varying an externally applied gate voltage, 
the linear conductance then typically
displays a sequence of peaks, which can be interpreted as 
Coulomb oscillations or resonant tunneling
peaks, depending on the parameter regime. 
The peak spacing is governed by the charging energy $E_c$ and 
the plasmon level spacing $\varepsilon$  on the dot 
(the latter coincides with the single-particle level spacing $E_s$ 
for the case of noninteracting electrons). 
We mention in passing that a similar double-barrier
experiment has been performed for semiconductor quantum wires, where in fact 
a TLL power-law temperature dependence  of the
peak conductance was reported.\cite{auslaender}
  In the nanotube experiments,\cite{Postma01,Bozovic01} 
two intramolecular barriers have been
 created within an individual metallic SWNT, 
and a power-law temperature
dependence of the conductance maximum  has been
 reported.\cite{Postma01}  However, the exponent turned out to be
inconsistent with expectations based on the sequential 
tunneling theory for a TLL; see below.
The experiments clearly indicate that
a detailed theoretical understanding of
transport through a double barrier in a TLL is required. 

Since the initial theoretical work on this topic,\cite{Kane92,Furusaki}
the double-barrier problem in a TLL has attracted a significant
amount of attention among 
theorists.\cite{Sassetti95,Furusaki98,Braggio00,Thorwart02,Nazarov03,Polyakov03,Komnik03,Hugle04,meden}
Those works show that the electron-electron interactions present in 1D
leads imply significant deviations from the conventional theory
for Fermi liquid leads.\cite{Kulik75,Beenakker91}  
This is most easily seen through the change of the temperature dependence of
the linear conductance peak height, which now
exhibits a typical TLL power law
$G_{\rm max}(T) \propto T^{\alpha}$, where $\alpha$ depends
on the interaction strength.  As is discussed in detail below, for a TLL,
the exponent $\alpha$  reveals the particular tunneling mechanism ruling
transport through the dot.
SWNTs possess an additional flavor 
degeneracy,\cite{egger,kane} 
which (together with the  spin degree of freedom)
can be captured by an effective four-channel TLL model. 
For an analysis of the linear conductance
on resonance, which is the focus of our 
work, however, a spinless single-channel TLL turns out to be
 sufficient.\cite{Kane92,Furusaki98,Hugle04} 
Within the standard bosonization method,\cite{sh02} 
the TLL is then characterized by a boson field $\vartheta(x)$ 
 with conjugate momentum $\Pi(x)$.  
Including a symmetric double barrier, composed
of short-range scattering centers at $\pm x_0/2$ of strength $U_{\rm imp}$,
the basic Hamiltonian  is\cite{Kane92,Furusaki}
\begin{equation} \label{hamg1}
H =  \frac{v_F}{2} \int dx  \left \{ \Pi^2 + \frac{1}{g^2}
(\partial_x \vartheta)^2 \right \} + H_{\rm imp}+
H_{\rm ext},
\end{equation}
where $v_F$ is the Fermi velocity,
\[
H_{\rm imp} =  U_{\rm imp} \sum_{\pm} \cos[\sqrt{4\pi} 
\vartheta(\pm x_0/2)]  ,
\]
 and we put $\hbar=k_B=1$ throughout this paper.
Asymmetric barriers can also be studied using our approach below,
but to keep the discussion as simple as possible, we restrict
ourselves to the symmetric case alone.
The electron-electron interaction
strength in the leads is measured in terms of the standard dimensionless
TLL parameter $g$, where $g=1$ for a Fermi gas and 
$g<1$ for repulsive interactions.\cite{sh02,Giamarchi04} 
We then have an intrinsic quantum dot within the TLL, with a 
plasmon level spacing $\varepsilon=E_s/g$, where $E_s=\pi v_F/x_0$
is the single-particle level spacing, and a
charging energy $E_c=E_s/g^2$. (Note that our convention for $E_c$ 
differs from the standard one.\cite{in92}) Finally,
the coupling to an applied bias voltage $V$ and to an external 
gate voltage $V_G$, 
acting onto the dot's electrons via a capacitance $C_{\rm G}$,
is encoded in\cite{Kane92,Furusaki}
\begin{equation} \label{driving}
H_{\rm ext}=-e (VN/2 +c V_G n ),
\end{equation}
where $c=C_G/(C+C_G)$ with the
island capacitance $C$.  In Eq.~(\ref{driving}), we used the notation
\begin{eqnarray}
N &=& \frac{1}{\sqrt{\pi}}\left[\vartheta \left(\frac{x_0}{2}\right)+\vartheta
\left(-\frac{x_0}{2}\right)\right],
\nonumber\\
n&=& \frac{1}{\sqrt{\pi}}\left[\vartheta \left(\frac{x_0}{2}\right)-\vartheta
\left(-\frac{x_0}{2}\right)\right]+n_0 ,
\end{eqnarray}
where $-eN$ gives the charge difference between the left and right
leads, $-en$ is the total electronic charge occupying the dot, 
and $n_0$ describes a possible offset charge.
Note that with our conventions, $N$ decreases (increases) when
electrons are transferred towards the right (left) lead. Likewise,
$n$ increases (decreases) for tunneling onto (out) of the dot. 
The Hamiltonian (\ref{hamg1}) has been studied in most of the previous works
on the subject and also forms the basis of our work.

The linear conductance through the dot
 is  periodic in $N_0=ceV_G/E_s$, and we can restrict ourselves
to one period, $0\leq N_0\leq 1$, where $G$ has
a resonance peak at $N_0=1/2$.
One can compute $G$ analytically for the non-interacting limit, $g=1$,
by refermionizing this model,\cite{Hugle04}
\begin{equation}\label{cond1}
\frac{G_{g=1}}{G_0} = \int
 \frac{dE}{4T \cosh^2(E/2T)} 
\frac{ w^2}{\cos^2(\pi[N_0+E/E_s]) + w^2 } \, ,
\end{equation}
with $G_0=e^2/h$ and 
$w = (4-\lambda^2)^2/[8\lambda(4+\lambda^2)]$, where
$\lambda = \pi U_{\rm imp}/D$ for the bandwidth $D$,  
which provides a high-energy cut-off.
For strong barriers, Eq.~(\ref{cond1})
leads to the standard Breit-Wigner resonant tunneling line shape
with linewidth $\Gamma_0= w E_s/\pi$. (For $g=1$,
the infinite-barrier limit is reached already for $\lambda=2$, where
the associated phase shift is in the unitary limit.)
Unfortunately, as the model (\ref{hamg1}) is not integrable for $g<1$,
exact solutions covering a wide parameter range of interest for this
transport problem are out of reach.
Analytical progress then generally has to rely on approximations.
Initial work \cite{Kane92,Furusaki,Sassetti95} pursued  perturbative
approaches, using the renormalization group, instanton methods or cumulant
expansions, both for strong and weak barriers.
Furusaki \cite{Furusaki98}  presented a detailed study of the
{\sl uncorrelated sequential tunneling} (UST)
regime, including also cotunneling contributions important away from the
resonance peak.  This regime allows for a {\sl master equation approach},
whose validity requires that the barriers are sufficiently strong and
that the temperature $T$ is sufficiently high.  
For $T$ well above $\varepsilon$ but still below $E_c$, 
the discrete nature of the plasmon modes on the dot is
not relevant while Coulomb blockade still exists. Then 
the corresponding results of Ref.~\onlinecite{Kane92}
apply. In particular, one obtains power-law scaling for the peak conductance
with $\alpha=2(1/g-1)$, implying that in this case
the double barrier effectively acts as a single
impurity. In what follows, we only discuss the case  $T<\varepsilon$.

Keeping only rates to lowest
order in $\Gamma_0$ from the TLL leads onto the island (dot),
which is equivalent to taking the standard UST mechanism,
the conductance is given by\cite{Furusaki98}
\begin{equation} \label{seq}
\frac{G_{\rm UST}}{G_0} =
\frac{\Gamma_0 (\pi T/D)^{-1+1/g} }{4 \Gamma(1/g) T \cosh(  \delta/2T)}
\left|\Gamma\left( \frac{1}{2g} +
  \frac{i\delta}{2\pi T} \right)\right|^2 ,
\end{equation}
where $\Gamma(x)$ denotes the Gamma function, and
$\delta = E_c |N_0-1/2|$ measures the distance from the resonance
peak.  The hybridization $\Gamma_0 =2\pi \rho_0 \Delta^2$
can be expressed in terms
of the 1D density of states, $\rho_0=1/(\pi v_F)$,
and the hopping amplitude $\Delta$ onto the dot,\cite{Weiss99} 
\begin{equation}\label{gamma0}
\Delta = \pi^{-1} \Gamma(1+1/g) \Gamma^{1/g}(1+g) (\pi U_{\rm imp}/ D)^{-1/g} D.
\end{equation}
The line shape (\ref{seq}) is rather close to the Fermi-liquid form
(\ref{cond1}) for large barrier heights and
evidently characterized by a linear $T$
dependence of the linewidth.  In the tails of a peak, the 
conductance (\ref{seq}) vanishes exponentially, but
then also  (elastic) cotunneling has to be 
included.\cite{Furusaki,Sassetti95,Furusaki98}
Note that the UST peak conductance in Eq.~(\ref{seq}) scales as
$G_{\rm max}\propto T^{-2+1/g}$, so that the power-law
exponent is  $\alpha_{\rm UST}=-2+1/g$.
Finally, at low temperatures, instead of sequential tunneling,
{\sl coherent resonant tunneling} is possible, characterized by
non-Lorentzian universal line shapes,\cite{Kane92}
where the linewidth scales as $T^{1-g}$.  In this paper, we 
only address the incoherent regime by employing a Markovian 
 master equation approach.  
Moreover, we focus on the temperature dependence of 
the peak conductance,  for which cotunneling is always a subleading
process.\cite{Furusaki98}
Therefore we neglect cotunneling throughout this work.
This is not a fundamental
restriction to our approach, but implies some technical simplifications.
On resonance, the basically only restriction for the validity
of our master equation approach is then given by the condition 
$G_{\rm max}\ll G_0$.\cite{Furusaki98}

Remarkably, the available experimental data obtained in SWNTs\cite{Postma01}
seem incompatible with Eq.~(\ref{seq}),
since the observed temperature dependence of the
conductance peak height $G_{\rm max}$ does not follow the UST scaling
but rather suggests a $T^{-3+2/g}$ power law.  The exponent 
$\alpha_{\rm CST}=-3+2/g$ has been
proposed to arise from a ``correlated sequential
tunneling'' (CST) mechanism.\cite{Thorwart02}
CST processes are sequential tunneling processes that cannot be
subdivided into two uncorrelated steps, yet still can be captured
in a master equation framework.  Such processes have been studied
previously in the chemical physics community, e.g., in the context of 
electron transfer reactions through an intermediate
bridge state.  In particular, Hu and Mukamel\cite{hu} have
treated the sequential tunneling regime for this problem using 
a very similar master equation theory.
Here we compute the sequential tunneling current through a
double barrier in a TLL, see Eq.~(\ref{hamg1}), beyond the lowest order
in $\Gamma_0$, implying modifications to Eq.~(\ref{seq}). 
We analyze to what extent such processes could indeed cause power-law scaling 
in $G_{\rm max}(T)$ with the CST exponent $\alpha_{\rm CST}$.

The remainder of the paper is structured as follows.
In Sec.~\ref{sec2} we introduce the general master equation approach, 
 and apply it to the regime of 
linear transport.  The transition rates entering the master equation
are evaluated in Sec.~\ref{sec3}, and in
 Sec.~\ref{sec4} explicit results for the temperature dependence
of the conductance peak, $G_{\rm max}(T)$, are presented.
 We conclude by discussing our results and 
their relation to other (experimental as well as theoretical)
work in Sec.~\ref{sec5}. 
Technical details concerning Sec.~\ref{sec3} can been found in the Appendix. 

\section{Markovian master equation}\label{sec2}

Let us start with the case of large tunneling barriers $U_{\rm imp}$,
which can be described within a dual version of Eq.~(\ref{hamg1}), see
Refs.~\onlinecite{Kane92,Furusaki,Furusaki98,Weiss99}. 
In this regime, the dynamics is dominated by tunneling events 
connecting minima of the periodic 
potential $H_{\rm imp}=2U_{\rm imp} \cos(\pi N) \cos(\pi n)$
 in the $(N,n)$-plane.\cite{Kane92} 
Such tunneling events induce a change $n\to n\pm 1$
 for tunneling onto/out of the dot, and $N\to N\mp 1$ for tunneling
towards the right/left. Hence transfer of one unit of
charge across the complete double barrier structure requires $N\to N\pm 2$.
The current through the double barrier is then given by
\begin{equation} \label{currentR}
I= \frac{e}{2}\langle \dot N \rangle,
\end{equation}
where the expectation value stands for a quantum-statistical average
with Hamiltonian (\ref{hamg1}), and Eq.~(\ref{currentR})
has to be evaluated in the stationary long-time limit.  
The discrete dynamics underlying Eq.~(\ref{currentR}) can be described
by a master equation\cite{kubo} for the probability
$P_N(n,t)$ of being in state $(N,n)$ at time $t$.

Master equations have previously been employed  for the
non interacting case,\cite{Kulik75,Beenakker91,in92} and for the TLL case
to lowest order in $\Gamma_0$.\cite{Furusaki98,Braggio00}
The rates entering the master equation can be extracted as
irreducible diagrams for the self-energy,\cite{Weiss99}
which to lowest order are simple golden rule rates.
These first-order UST contributions to the transition rates 
imply $N\to N\pm 1$ jumps.  Below
we will systematically 
take into account transition rates up to second order in $\Gamma_0$.
It turns out that the second-order contributions
to those rates are plagued by nontrivial
divergences, which require
a resummation of higher-order diagrams.  This resummation is done below
by employing the Weisskopf-Wigner approximation.\cite{Wigner30}
On the one hand, this procedure yields a direct CST process,
$N\to N\pm2$. On the other hand, additional (indirect) CST 
contributions to the transition rates for $N\to N\pm 1$ arise. 
Apart from the need of a regularization scheme,
the same situation is encountered in the theory of bridged electron transfer
reactions.\cite{hu}

The master equation for $P_N(n,t)$ then has the general structure\cite{kubo}
\begin{eqnarray} \label{mastereq}
\lefteqn{\dot{P}_N(n,t)  =  - \gamma(n) P_N(n,t)}  \\ 
& & + \Gamma_R^f(n+1) P_{N+1}(n+1,t)
+ \Gamma_L^b(n+1) P_{N-1}(n+1,t) \nonumber \\
& & + \Gamma_L^f(n-1) P_{N+1}(n-1,t) + \Gamma_R^b(n-1) P_{N-1}(n-1,t)
\nonumber \\
& & +\Gamma_{CST}^f(n)P_{N+2}(n,t) + \Gamma_{CST}^b(n)P_{N-2}(n,t)
\nonumber \\ \nonumber
& & -\Gamma_{CST}^f(n)P_{N}(n,t) - \Gamma_{CST}^b(n)P_{N}(n,t) ,
\end{eqnarray}
where $\Gamma_{L/R}^{f/b}(n)$ denotes the forward/backward rate for 
a transition over the left/right barrier, having {\sl started}\
with $n$ electrons on the dot. In addition, $\Gamma_{CST}^{f/b}(n)$ 
denotes the forward/backward rate for a direct CST transition 
$N\to N\pm 2$.  Moreover, we use
\begin{equation} \label{linewidth}
\gamma (n) = \Gamma_R^f(n)+\Gamma_L^f(n)+\Gamma_R^b(n)+\Gamma_L^b(n),
\end{equation}
which is related to the linewidth of the state $(N,n)$. 
It is useful to also introduce the probability 
\begin{equation}\label{pnt}
p(n,t)= \sum_{N=-\infty}^{+\infty}P_N(n,t)
\end{equation}
for the dot being occupied with $n$
electrons at time $t$.
In order to calculate the current $I$, we 
insert Eq.~(\ref{mastereq}) into Eq.~(\ref{currentR}), which yields 
after some algebra 
\begin{eqnarray}
I & = & - \frac{e}{2} \sum_{n=-\infty}^{+\infty} 
\Bigl[
\Gamma_R^f(n)+\Gamma_L^f(n)-\Gamma_R^b(n)-\Gamma_L^b(n) \nonumber \\
 &+ &  2 \Gamma_{CST}^f(n)-2\Gamma_{CST}^b(n)
\Bigr] p(n,t\to \infty). \label{current3}
\end{eqnarray}
Summing both sides of Eq.~(\ref{mastereq}) over $N$, we obtain
a master equation for $p(n,t)$ directly,
\begin{eqnarray} \label{mesmall}
\dot{p}(n,t) & = & - \gamma(n) p(n,t)  \\ 
&+ &  \left[ \Gamma_L^b(n+1) + \Gamma_R^f(n+1) \right] p(n+1,t) \nonumber \\
\nonumber &+& \left[ \Gamma_L^f(n-1) + \Gamma_R^b(n-1) \right] p(n-1,t)  .
\end{eqnarray}
The stationary solution $p(n)=p(n,t\to \infty)$  
 follows by requiring $\dot{p}(n)=0$, which yields
the detailed balance relation
\begin{equation} \label{detbal}
\frac{p(n)}{p(n+1)} = \frac{\Gamma_L^b(n+1) + 
\Gamma_R^f(n+1) }{\Gamma_L^f(n) + \Gamma_R^b(n)}.
\end{equation}
Taking into account conservation of probability, 
$\sum_{n=-\infty}^{\infty}p(n)=1$,
this relation can be solved recursively.
We note that the direct CST rates $\Gamma_{CST}^{f/b}(n)$ 
do not appear in Eq.~(\ref{mesmall}), since they do 
not alter the net population of the 
dot. However, they do appear in the current (\ref{current3}).

Let us now focus on the linear transport regime and sufficiently low
temperatures,  $eV, T\ll E_c$, where at most two charge states 
$n$ are allowed on the dot due to charging effects.\cite{in92} 
 Without loss of generality, we may choose
 $n=0$ and $n=-1$ to label those states.
In the linear transport regime, we can disregard the rates 
$\Gamma_L^f(0), \Gamma_R^f(-1), \Gamma_R^b(0)$, and $\Gamma_L^b(-1)$,
 since they involve energetically forbidden states with 
dot occupation numbers $n=+1$ or $n=-2$. The  
 recursive solution of the detailed balance relation 
(\ref{detbal}) then yields 
\begin{eqnarray}
\label{probs}
p(0) & = & \frac{\Gamma_L^f(-1) + \Gamma_R^b(-1)}
{\Gamma_L^b(0) + \Gamma_R^f(0) + \Gamma_L^f(-1) + \Gamma_R^b(-1)}\, ,\\
 \nonumber 
p(-1) & = & \frac{\Gamma_L^b(0) + \Gamma_R^f(0)}
{ \Gamma_L^b(0) + \Gamma_R^f(0)+\Gamma_L^f(-1) + \Gamma_R^b(-1) }.
\end{eqnarray}

Combining Eqs.~(\ref{probs}) and
(\ref{current3}), we obtain
$I= I_1 + I_2$, with the standard
 contribution\cite{Kulik75,Beenakker91,in92,Furusaki98,Braggio00}
\begin{equation}\label{currentx1}
I_1 = -e \frac{\Gamma_R^f(0) \Gamma_L^f(-1)-\Gamma_R^b(-1) \Gamma_L^b(0)}
{\Gamma_R^f(0)+ \Gamma_L^f(-1)+\Gamma_R^b(-1) +\Gamma_L^b(0)} ,
\end{equation}
and an additional term caused by direct CST rates,
\begin{widetext}
\begin{equation} \label{currentx2}
I_2 = -e \frac{ \left[ \Gamma_{CST}^f(0) - \Gamma_{CST}^b(0) \right]
\left[ \Gamma_L^f(-1)+\Gamma_R^b(-1) \right] +
\left[ \Gamma_{CST}^f(-1) - \Gamma_{CST}^b(-1) \right] \left[
\Gamma_L^b(0)+\Gamma_R^f(0) \right]
}{\Gamma_R^f(0)+ \Gamma_L^f(-1)+\Gamma_R^b(-1) +\Gamma_L^b(0)} .
\end{equation}
\end{widetext}
So far we have discussed a general procedure to determine the
current in the linear regime. 
To make progress,  the transition rates entering the above equations
must be computed for the Hamiltonian in Eq.~(\ref{hamg1}).

\section{Transition rates} \label{sec3}

\begin{figure}
\begin{center}
\epsfig{figure=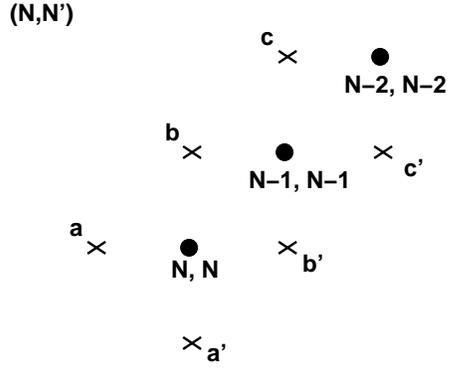,width=60mm,angle=0,keepaspectratio=true}
\caption{Relevant part of the $(N,N')$-plane of the RDM (schematic).
Diagonal states are indicated by filled circles, off-diagonal states are 
marked by crosses. We use the shorthand notation $a= (N,N+1) , 
 b=( N-1,N)$,  and $c=(N-2,N-1)$, and complex conjugate states are
indicated by the prime. 
For the irreducible $\Gamma_0^2$ contribution to the $N\rightarrow N-1$ rate,
 we have four jumps. One starts
 from $(N,N)$ and ends in $(N-1,N-1)$, visiting an intermediate 
diagonal state after every second jump.
 \label{fig.plan}}
 \end{center}
\end{figure}

We now systematically compute all rates entering 
Eqs.~(\ref{currentx1}) and (\ref{currentx2}).  Rates of 
order higher than $\Gamma_0^2$ 
will be included approximately within the Weisskopf-Wigner scheme.
For consistency, besides the direct CST rates,
 it is also mandatory to include all indirect CST contributions 
to the rates $\Gamma_{\lambda}^{\nu}(n)$
with $\lambda=L/R$ and $\nu=f/b$.
All these transition rates  can be extracted as the 
irreducible components of an exact perturbation series expression
for the probability distribution $P_N(n,t)$. The latter corresponds
to the diagonal element of the reduced density matrix (RDM), which in turn
allows for a path-integral representation.\cite{Weiss99} 
 Tracing out the Gaussian TLL modes away from the barriers at
 $x=\pm x_0/2$, one obtains an effective action which is 
 equivalent to the action of a quantum Brownian particle  hopping 
in the $(N,n)$-plane.\cite{Kane92,Sassetti95}   Then a path can be
visualized in the $(N,N')$-plane of the RDM, see
Fig.~\ref{fig.plan}, with a corresponding
dynamics in the $(n,n')$-plane.
Alternatively, the rate expressions  given below can 
also be derived using the projection operator
formalism.\cite{hu,kubo,zwanzig}
The rates  $\Gamma_{\lambda}^{\nu}(n)$ for a transition $N\to N\pm 1$
then have contributions of first and at least second order in $\Gamma_0$,
\begin{equation} \label{gam1}
\Gamma_{\lambda}^{\nu}(n) = \Gamma_{\lambda}^{\nu,(1)}(n) + 
\Gamma_{\lambda}^{\nu,(2)}(n),
\end{equation}
where we keep only the $n=0,-1$ states.

\subsection{First-order rates}\label{sec31}

\begin{figure}
\begin{center}
\epsfig{figure=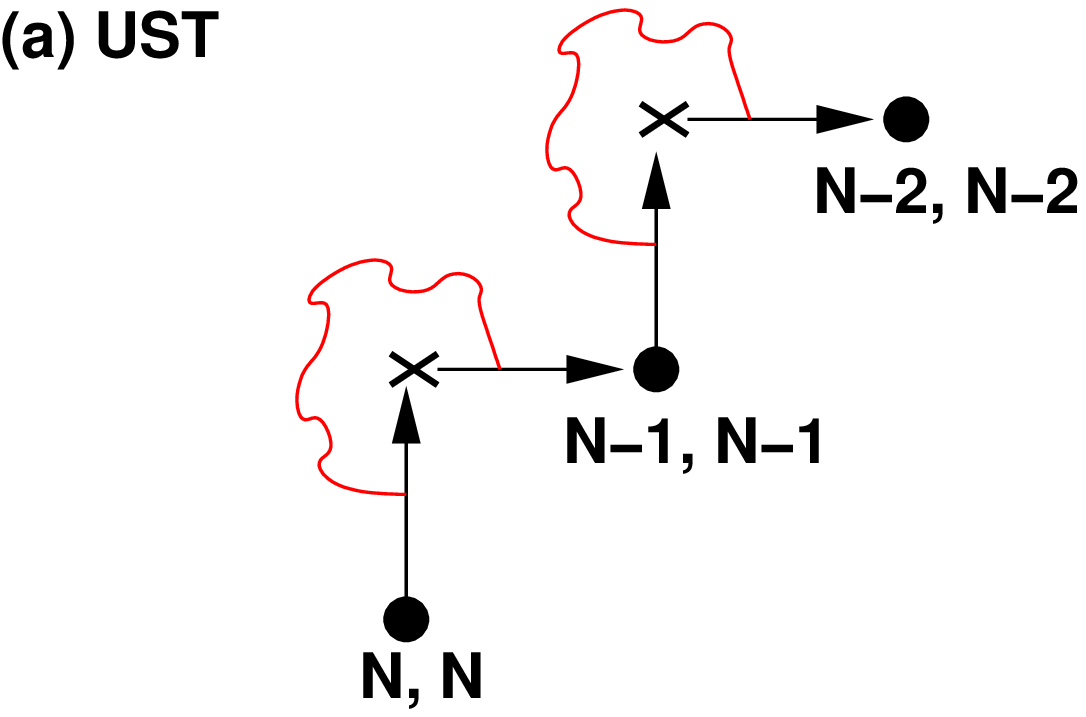,width=60mm,angle=0,keepaspectratio=true}
\epsfig{figure=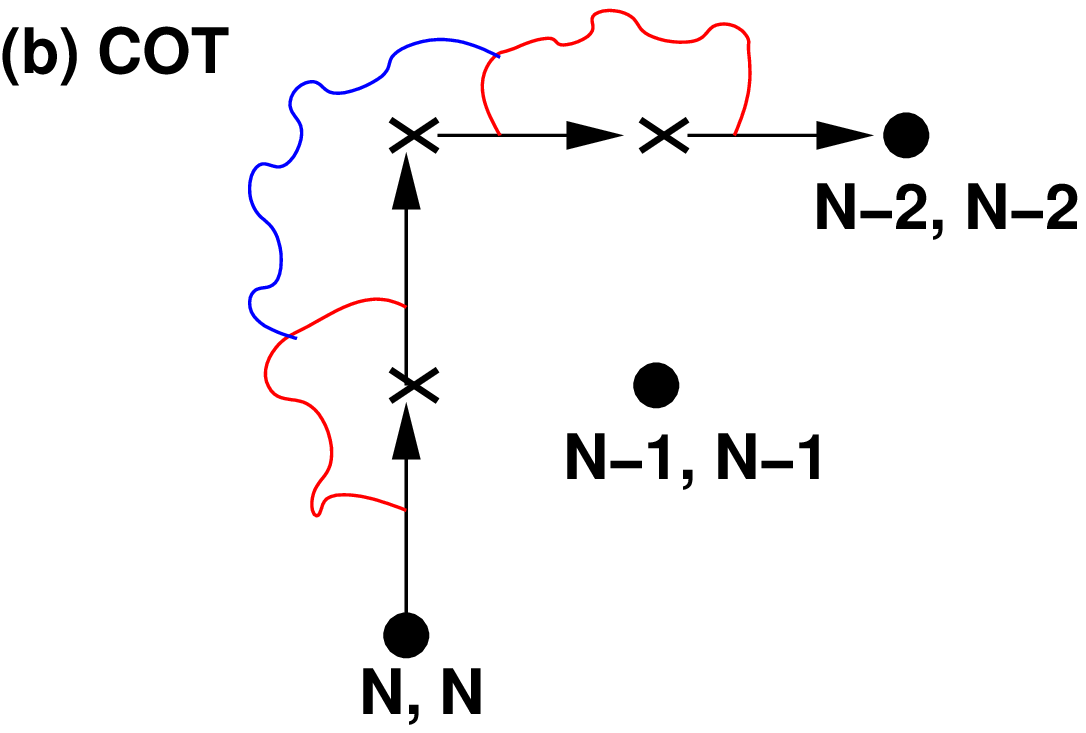,width=60mm,angle=0,keepaspectratio=true}
\caption{(Color online) Paths in the RDM for (a) 
uncorrelated sequential tunneling (UST), and
(b) cotunneling (COT). Wiggled lines schematically
indicate ``bath-induced'' correlations 
for first (red)- and second (blue)-order 
transitions in $\Gamma_0$.
Diagram (a) involves two irreducible golden rule transition rates,
i.e., there are no correlations across the intermediate diagonal state.
Diagram (b) is not considered in what follows since we study a
conductance peak.  \label{fig.ust} }
 \end{center}
\end{figure}

The first-order contribution is 
well-known,\cite{Kane92,Furusaki,Sassetti95,Furusaki98,Weiss99}
and schematically depicted in Fig.~\ref{fig.ust}(a).  
There are two independent (uncorrelated) steps, one from the left
TLL ``lead'' onto the island, and another to the right TLL ``lead''.
Using the hopping matrix element $\Delta$ in Eq.~(\ref{gamma0}), 
these steps individually correspond to irreducible golden rule rates, 
\begin{equation}\label{lowestorder}
 \Gamma_{\lambda}^{\nu,(1)}(n) = \frac{\Delta^2}{2}
 \mbox{\rm Re} \int_0^{\infty} dt
 \exp[i E_{\lambda \nu}(n)t - W_{\Sigma}(t) ] ,
\end{equation}
where $W_\Sigma(t)=W_+(t)+W_-(t)$, with the
correlation functions\cite{Sassetti95}
\begin{eqnarray} \label{corrfunc}
 W_{\pm}(t) &=& \int_{0}^{\infty}d\omega \frac{J_{\pm}(\omega)}{\omega^2} 
 \{ [1-\cos(\omega t)] \\  \nonumber
 & & \times\coth(\omega/2T) + i\sin(\omega t)\}.
\end{eqnarray}
The spectral densities follow as 
\begin{equation} \label{spectral}
 J_\pm(\omega)=\frac{\omega e^{-\omega/D}}{2g}
 \left[1+2\varepsilon\sum_{m=1}^{\infty}
 \delta (\omega -{\cal M}_{\pm}(m)\varepsilon)\right],
\end{equation}
 where ${\cal M}_+(m)=2m-1$, ${\cal M}_-(m)=2m$, and 
 $\varepsilon=E_s/g=\pi v_F/ (gx_0)$ is the plasmon level 
spacing on the dot. 
The $\delta$-peaks are a result of the finite level spacing on the dot,
while the first part in Eq.~(\ref{spectral}) yields a standard
Ohmic spectral density.
The bandwidth $D$ is taken as smooth (exponential) 
ultraviolet cutoff for the model (\ref{hamg1}).  These correlation functions
arise in the process of integrating out the TLL modes away from
the barrier, see Ref.~\onlinecite{Weiss99} for a detailed review of this
procedure. 
Finally,
the energies $E_{\lambda,\nu}(n)$ appearing in
Eq.~(\ref{lowestorder}) are defined as
\begin{eqnarray}
E_{Rf}(n+1)&=&-E_{Rb}(n)=\mu(n+1)-eV/2,\nonumber\\
E_{ Lb}(n+1)&=&-E_{Lf}(n)=\mu(n+1)+eV/2 , \label{bias}
\end{eqnarray}
with the electrochemical potential 
\[
\mu(n+1)= 2 E_c(n-n_0-ecV_G/2E_c +1/2).
\]
The correlation function $W_{\Sigma}(t)=W_++W_-=W_{\rm Ohm}+W_{\rm dot}$ 
can be decomposed into two different parts,\cite{Sassetti95,Braggio00}
namely an Ohmic part $W_{\rm Ohm} (t)$ and an oscillatory part
$W_{\rm dot} (t)$. Here the first contribution comes from the smooth 
part in $J_{\pm}(\omega)$, and takes the standard form
\begin{eqnarray}\nonumber
W_{\rm Ohm}(t) &=& \frac{1}{g}\ln[(D/T)\sinh|\pi T t|] + i(\pi/g)
 {\rm sgn}( t)\\
&=& S_{\rm Ohm}(t) + i R_{\rm Ohm}(t). \label{ohmic}
\end{eqnarray}
At low temperatures, $T\ll \varepsilon$, the dot correlation function
is given by its $T=0$ limiting form
\begin{equation}\label{Wdot}
W_{\rm dot}(t)=\frac{1}{g}\ln\left(
\frac{1-e^{-(\varepsilon/D +i\varepsilon t)}}{1
 -e^{-\varepsilon/D}} \right),
\end{equation}
up to exponentially small corrections in $y=e^{-\varepsilon/T}$. 
This function arises due to the finite level spacing,
 and is periodic in $t$
 with period $\tau_\varepsilon=2\pi/\varepsilon$.
 It can therefore be expanded in a Fourier series,\cite{Braggio00} leading to 
\begin{equation}
 \label{ustrate}
 \Gamma_{\lambda}^{\nu,(1)}(n) = \sum_{p=-\infty}^{+\infty} d_p(\varepsilon) \;
 \Gamma_{\rm Ohm} (E_{\lambda \nu}(n) -p \varepsilon) . 
\end{equation}
The Fourier coefficients $d_p(\varepsilon)$ can be found in  the
Appendix, and from Eq.~(\ref{ohmic}), one gets\cite{Weiss99}
\begin{equation}\label{ohmicrate}
 \Gamma_{\rm Ohm}(E) = \frac{\Delta^2}{4 D}
 \frac{e^{E/2T}}{\Gamma(1/g)} \left(\frac{D}{2 \pi T}\right)^{1-1/g}
 \left|\Gamma \left( \frac{1}{2g}+  \frac{iE}{2 \pi T}\right)\right|^2
 \, ,
\end{equation}
which in turn directly leads to Eq.~(\ref{seq}). We note that
the first-order forward/backward rates (\ref{lowestorder})  fulfill 
a detailed balance relation
\begin{equation}\label{detbalance}
\Gamma_{\lambda}^{b,(1)}(n-1) = 
e^{-E_{\lambda f}(n)/T} \Gamma_{\lambda}^{f,(1)}(n) ,
\end{equation}
which formally follows from the reflection property 
$W_\pm(-t)= W_\pm^*(t)=W_\pm(t-i/T)$ of the above
correlation functions.

\subsection{Indirect CST rates}\label{sec32}

\begin{table}
\begin{ruledtabular}
\begin{tabular}{c l}
$k$ & Indirect transitions $N \rightarrow N-1$: \\
\hline
1 & $ N,N \rightarrow a \rightarrow N,N \rightarrow b \rightarrow N-1,N-1 $\\
2 & $ N,N \rightarrow a \rightarrow N,N \rightarrow b' \rightarrow N-1,N-1 $\\
3 & $ N,N \rightarrow b \rightarrow N,N \rightarrow b \rightarrow N-1,N-1 $\\
4 & $ N,N \rightarrow b \rightarrow N,N \rightarrow b' \rightarrow N-1,N-1 $\\
5 & $ N,N \rightarrow b \rightarrow N-1,N-1 \rightarrow b \rightarrow N-1,N-1 $\\
6 & $ N,N \rightarrow b \rightarrow N-1,N-1 \rightarrow b' \rightarrow N-1,N-1 $\\
7 & $ N,N \rightarrow b \rightarrow N-1,N-1 \rightarrow c \rightarrow N-1,N-1 $\\
8 & $ N,N \rightarrow b \rightarrow N-1,N-1 \rightarrow c' \rightarrow N-1,N-1 $\\
\hline
$\#$ & Direct transitions $N \rightarrow N-2$: \\
\hline
CST1 & $N,N  \rightarrow   b \rightarrow N-1, N-1 \rightarrow c \rightarrow N-2,N-2 $ \\
CST2 & $N,N  \rightarrow   b \rightarrow N-1, N-1 \rightarrow c' \rightarrow N-2,N-2 $ \\
\end{tabular}
\end{ruledtabular}
\caption{All transition processes of order 
$\Gamma^2_0$ contributing to the master equation.
The off-diagonal states $a,b$ and $c$
are specified in Fig.~\ref{fig.plan}.
\label{tab.paths}}
\end{table}

\begin{figure}
\begin{center}
\epsfig{figure=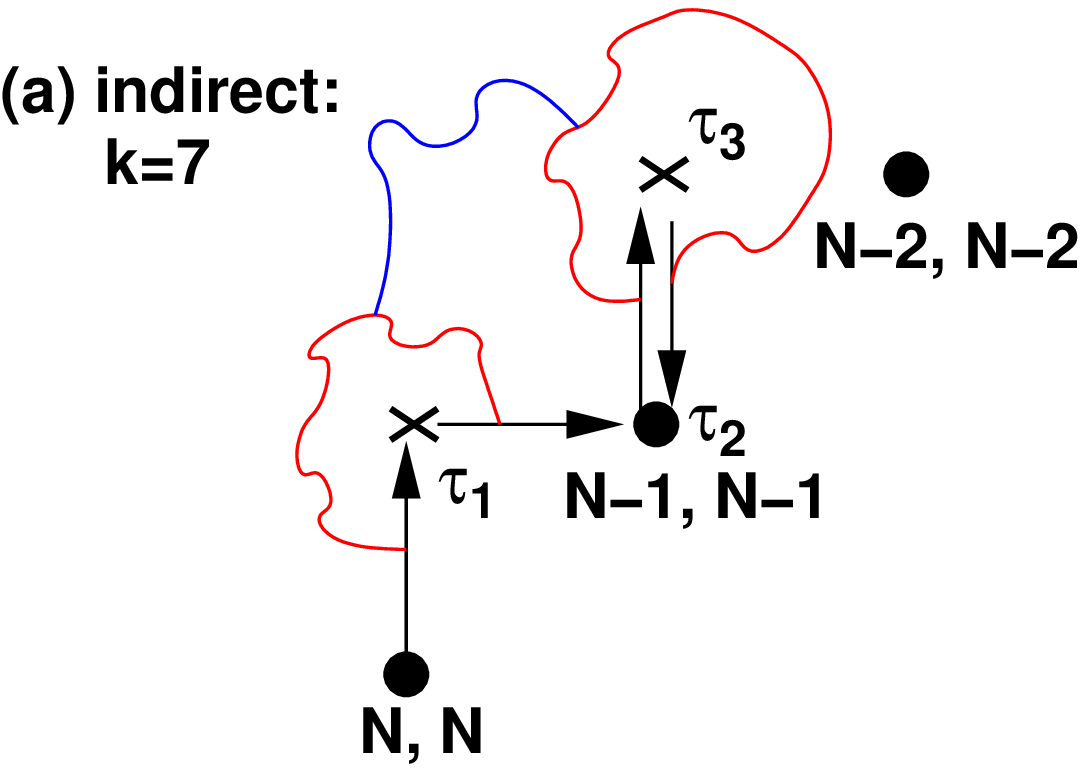,width=60mm,angle=0,keepaspectratio=true}
\epsfig{figure=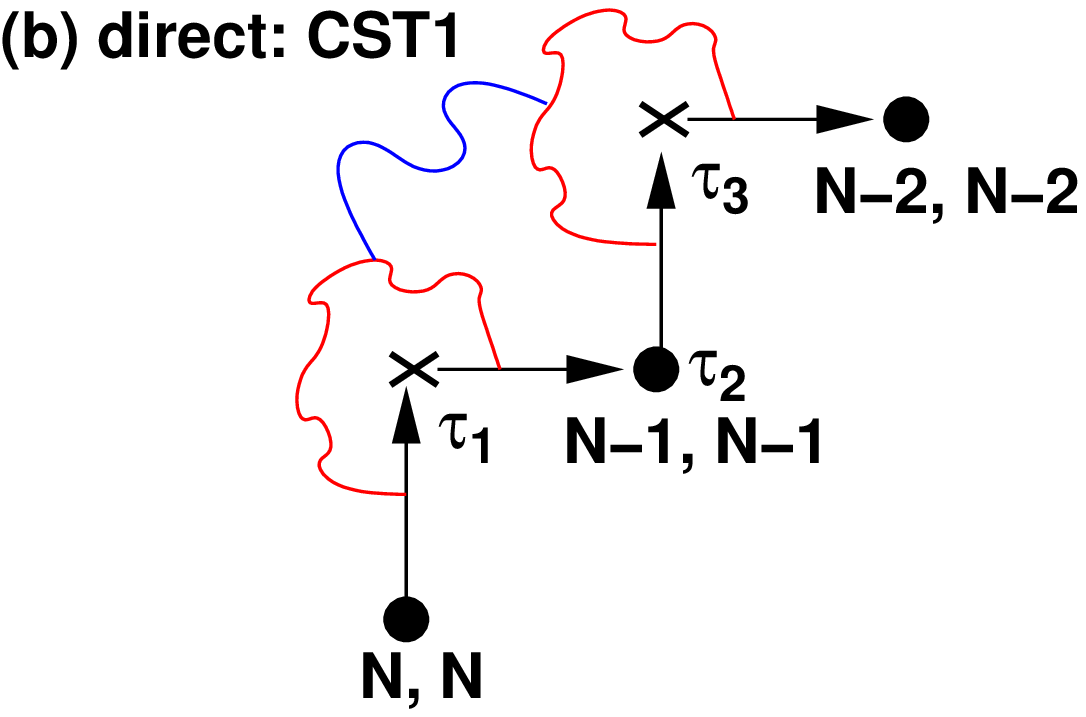,width=60mm,angle=0,keepaspectratio=true}
\caption{(Color online) Examples for CST
diagrams of order $\Gamma_0^2 \propto \Delta^4$: (a) the ``indirect'' diagram
$k=7$ in Tables \ref{tab.paths} and \ref{tab.diag} and  (b) the 
``direct'' diagram CST1.  \label{fig.cst} } 
\end{center}
\end{figure}

The $\Gamma_0^2$ contributions to the rate  for $N\to N-1$,
which we shall call ``indirect transitions'', require a careful counting
of all possible transitions 
consisting of four jumps in the $(N,N')$-plane, see 
Figure \ref{fig.plan} and Table \ref{tab.paths}. 
There are eight irreducible diagrams for the forward rate,
plus their complex conjugates, 
which can however be included by taking twice the real part.
Similarly, there are eight diagrams for the backward rate.
Each irreducible second-order diagram then gives a triple integral 
over three times $\tau_1, \tau_2$ and
$\tau_3$, which represent the time 
spent in the corresponding state of the reduced density matrix
 (RDM).\cite{Sassetti95,Weiss99}
To give a concrete example, the diagram denoted by $k=7$ in 
Table \ref{tab.paths} is drawn schematically in Fig.~\ref{fig.cst}(a). 
In particular, since we neglect cotunneling processes,
after two jumps we are always in a diagonal state of the RDM,
amounting  to a real (as opposed to virtual) occupation of the
corresponding state. Therefore $\tau_2$  in those expressions
will always have the meaning of the 
time spent in the respective intermediate diagonal state.
Similarly, for the ``direct'' CST rate with $N\to N-2$, there are two such
triple-integral contributions (plus their complex conjugate diagrams). 
At this point, we note that we label the direct transitions as CST1/2 and
the indirect transitions with the index $k=1, \dots, 8$, although in principle 
{\em all\/} second-order processes taken into account here
are correlated sequential tunneling processes. 

An important property of the irreducible second-order diagrams
is that for finite plasmon level spacing $\varepsilon$,
each of them contains a divergence.
 This is not a trivial divergence 
as we work with irreducible diagrams.
Formally, this infrared divergence comes from the $\tau_2$ integration that
extends from $0$ to $\infty$ but has an integrand which is ultimately periodic
in $\tau_2$. 
This implies that one has to effectively include higher-order diagrams,
 which is of course impossible to achieve in an exact way.
Here we use the Weisskopf-Wigner approximation\cite{Wigner30} 
to regularize the second-order diagrams.  Physically,
the intermediate state has a 
finite lifetime linked to the linewidth parameter $\gamma(n)$ in
Eq.~(\ref{linewidth}).  We then introduce a 
factor $e^{-\gamma(n) \tau_2}$ for the $\tau_2$-integrations, 
where the linewidth $\gamma(n)$ has to be computed self-consistently 
by  requiring that Eq.~(\ref{linewidth}) holds. 
This linewidth $\gamma(n)$ then
enters the master equations (\ref{mastereq}) and (\ref{mesmall}).

\begin{table*}
\begin{ruledtabular}
\begin{tabular}{l}
Forward $N\rightarrow N-1$ rates of (at least) order $\Gamma_0^2$
 over the right barrier: \\
\hline
$\Gamma_{R,1}^{f,(2)}(n) = -2 \frac{\Delta^4}{16} \mbox{\rm Re } \int_0^{\infty} d\tau_i \,$ \\
$ \mbox{\hspace{1ex}} \times e^{i [-E_{Lb}(n) \tau_1 +  E_{Rf}(n) \tau_3 ]}
e^{-W_{\Sigma}^*(\tau_1)-W_{\Sigma}(\tau_3)} \left[
e^{W_{\Delta}^*(\tau_1+\tau_2)-W_{\Delta}^*(\tau_2)-W_{\Delta}^*(\tau_1+\tau_2+\tau_3)
+ W_{\Delta}^*(\tau_2+\tau_3)}-1\right] e^{-\gamma(n) \tau_2}$ \\[3mm]
$\Gamma_{R,2}^{f,(2)}(n) =-2 \frac{\Delta^4}{16} \mbox{\rm Re } \int_0^{\infty} d\tau_i \,$ \\
$ \mbox{\hspace{1ex}} \times
e^{i [-E_{Lb}(n) \tau_1 -  E_{Rf}(n) \tau_3 ]}
e^{-W_{\Sigma}^*(\tau_1)-W_{\Sigma}^*(\tau_3)} \left[
e^{-W_{\Delta}^*(\tau_1+\tau_2)+W_{\Delta}^*(\tau_2)+W_{\Delta}^*(\tau_1+\tau_2+\tau_3)
- W_{\Delta}^*(\tau_2+\tau_3)}-1\right] e^{-\gamma(n) \tau_2} $\\[3mm]
$\Gamma_{R,3}^{f,(2)}(n) =-2 \frac{\Delta^4}{16} \mbox{\rm Re } \int_0^{\infty} d\tau_i \,$ \\
$ \mbox{\hspace{1ex}} \times
e^{i [E_{R f} (n) \tau_1  +  E_{R f} (n) \tau_3 ]}
e^{-W_{\Sigma}(\tau_1)-W_{\Sigma}(\tau_3)} \left[
e^{W_{\Sigma}(\tau_1+\tau_2)-W_{\Sigma}(\tau_2)-W_{\Sigma}(\tau_1+\tau_2+\tau_3)
+ W_{\Sigma}(\tau_2+\tau_3)}-1\right] e^{-\gamma(n) \tau_2} $\\[3mm]
$\Gamma_{R,4}^{f,(2)}(n) =-2 \frac{\Delta^4}{16} \mbox{\rm Re } \int_0^{\infty} d\tau_i \,$ \\
$ \mbox{\hspace{1ex}} \times
e^{i [E_{R f}(n)\tau_1 - E_{R f} (n) \tau_3]}
e^{-W_{\Sigma}(\tau_1)-W_{\Sigma}^*(\tau_3)} \left[
e^{-W_{\Sigma}(\tau_1+\tau_2)+W_{\Sigma}(\tau_2)+W_{\Sigma}(\tau_1+\tau_2+\tau_3)
- W_{\Sigma}(\tau_2+\tau_3)}-1\right] e^{-\gamma(n) \tau_2} $\\[3mm]
$\Gamma_{R,5}^{f,(2)}(n) =-2 \frac{\Delta^4}{16} \mbox{\rm Re } \int_0^{\infty} d\tau_i \,$ \\
$ \mbox{\hspace{1ex}} \times
e^{i [E_{R f}(n) \tau_1 - E_{R b} (n-1) \tau_3 ]}
e^{-W_{\Sigma}(\tau_1)-W_{\Sigma}^*(\tau_3)} \left[
e^{W_{\Sigma}(\tau_1+\tau_2)-W_{\Sigma}^*(\tau_2)-W_{\Sigma}(\tau_1+\tau_2+\tau_3)
+ W_{\Sigma}^*(\tau_2+\tau_3)}-1\right] e^{-\gamma(n) \tau_2} $\\[3mm]
$\Gamma_{R,6}^{f,(2)}(n) =-2 \frac{\Delta^4}{16} \mbox{\rm Re } \int_0^{\infty} d\tau_i \,$ \\
$ \mbox{\hspace{1ex}} \times
e^{i [E_{R f}(n) \tau_1 + E_{R b} (n-1) \tau_3 ]}
e^{-W_{\Sigma}(\tau_1)-W_{\Sigma}(\tau_3)} \left[
e^{-W_{\Sigma}(\tau_1+\tau_2)+W_{\Sigma}^*(\tau_2)+W_{\Sigma}(\tau_1+\tau_2+\tau_3)
- W_{\Sigma}^*(\tau_2+\tau_3)}-1\right] e^{-\gamma(n) \tau_2} $\\[3mm]
$\Gamma_{R,7}^{f,(2)}(n) =-2 \frac{\Delta^4}{16} \mbox{\rm Re } \int_0^{\infty} d\tau_i \,$ \\
$ \mbox{\hspace{1ex}} \times
e^{i [E_{R f} (n) \tau_1 + E_{Lf}(n-1) \tau_3 ]}
e^{-W_{\Sigma}(\tau_1)-W_{\Sigma}(\tau_3)} \left[
e^{W_{\Delta}(\tau_1+\tau_2)-W_{\Delta}^*(\tau_2)-W_{\Delta}(\tau_1+\tau_2+\tau_3)
+ W_{\Delta}^*(\tau_2+\tau_3)}-1\right] e^{-\gamma(n) \tau_2} $\\[3mm]
$\Gamma_{R,8}^{f,(2)}(n) =-2 \frac{\Delta^4}{16} \mbox{\rm Re } \int_0^{\infty} d\tau_i \,$ \\
$ \mbox{\hspace{1ex}} \times
e^{i [E_{R f} (n) \tau_1 - E_{Lf}(n-1) \tau_3 ]}
e^{-W_{\Sigma}(\tau_1)-W_{\Sigma}^*(\tau_3)} \left[
e^{-W_{\Delta}(\tau_1+\tau_2)+W_{\Delta}^*(\tau_2)+W_{\Delta}(\tau_1+\tau_2+\tau_3)
- W_{\Delta}^*(\tau_2+\tau_3)}-1\right] e^{-\gamma(n) \tau_2} $\\[3mm]
\end{tabular}
\end{ruledtabular}
\caption{The 8 irreducible rate expressions of at least order $\Gamma_0^2$,
 corresponding to
the forward-rate diagrams through the right barrier, $\Gamma_{R,k}^{f,(2)}$, 
see Eq.~(\ref{gam2}) and Table \ref{tab.paths}.
The corresponding rates through the left barrier,
$\Gamma_{L,i}^{f,(2)}(n-1)$, follow by
substituting $E_{Rf/Lb}(n)
\to E_{Lf/Rb}(n-1)$ and   $E_{Lf/Rb}(n-1)\to E_{Rf/Lb}(n)$.
 The backward rates $\Gamma_{R,i}^{b,(2)}(n-1)$ and
 $\Gamma_{L,i}^{b,(2)}(n)$ can be obtained from the forward rates
 $\Gamma_{R,i}^{f,(2)}(n)$ and $\Gamma_{L,i}^{f,(2)}(n-1)$ using
the substitutions $E_{Lb/Rf}(n) \to E_{Lf/Rb}(n-1)$ and
 $E_{Lf/Rb}(n-1) \to E_{Lb/Rf}(n)$, respectively.
\label{tab.diag}}
\end{table*}

The eight regularized rates $\Gamma_{\lambda,k}^{\nu,(2)}$ are given explicitly in
Table \ref{tab.diag}. At low temperatures, again the $T=0$ form
\begin{equation}\label{Wdelta}
W_{\Delta}(t)=W_+(t)-W_-(t) = -\frac{1}{g}\ln\left(
\frac{1+e^{-(\varepsilon/D +i\varepsilon t)}}{1
 +e^{-\varepsilon/D}} \right)
\end{equation}
holds up to exponentially small corrections
of order $y=e^{-\varepsilon/T}$, and
\begin{equation} \label{gam2}
\Gamma_{\lambda}^{\nu,(2)} = \sum_{k=1}^8 \Gamma_{\lambda,k}^{\nu,(2)}.
\end{equation}
This rate depends on $\gamma(n)$, which needs to be determined 
self-consistently. Note that $W_{\Delta}(t)$ is also periodic in $t$
 with period $\tau_\varepsilon$.

{}From now on, we always focus on the resonance peak, where all 
energies can be put to $E_{\lambda\nu}(n)=0$.
In addition, since on resonance $\mu(-1)=\mu(0)$, we also have
$\gamma(-1)=\gamma(0) = \gamma$.
The calculation of $\Gamma_{\lambda,k}^{\nu,(2)}$ is
comparatively simple for the diagrams  $k= 1,2,7,8$, since 
only the periodic correlation function $W_\Delta$
appears in the respective bracketed terms, which allows for
a Fourier series expansion.  Then 
 the triple integrals factorize, and the quickly converging
Fourier sums can easily be performed numerically.
For the benefit of the interested reader, 
in order to illustrate this procedure,
the evaluation of diagrams of this first class  
is discussed in full detail in the Appendix.
The remaining diagrams of the second class ($k=3,4,5,6$) are more difficult to handle,
since the bracketed terms now involve the correlation function $W_{\Sigma}$.
We have therefore evaluated the respective triple integrals 
numerically as a function of the linewidth $\gamma$.  This can be
done either via direct numerical quadrature (trapezoidal rule),
or using Monte Carlo integration \cite{numericalrecipes}.  The latter approach
is more suitable for large $\gamma$, where stochastic error bars can
be made arbitrarily small with only modest computational effort.
Fortunately, the $\gamma$-dependence of the four
diagrams of the second class ($k=3,4,5,6$)  
turns out to be identical to the one of the first class, which 
is given by Eq.~(\ref{fingam}) below, see also 
Eq.~(\ref{funcdep}) in the Appendix.
A fit of the numerical results for several $\gamma$ then
allows to accurately extract the parameters
$A_{g,k},B_{g,k}$ and $C_{g,k}$ for those diagrams as well.
Finally, we summarize {\sl all indirect CST contributions\/} as 
\begin{equation} \label{fingam}
\Gamma_{\lambda}^{\nu,(2)} = \frac{\Delta^4}{\varepsilon^3} \left(
-\frac{\varepsilon}{\gamma}A_g +\frac{\gamma}{\varepsilon} B_g +
C_g \right),
\end{equation}
where dimensionless parameters $A_g, B_g,$ and $C_g$ 
follow by summing over the 
respective values $A_{g,k}, B_{g,k}, C_{g,k}$ for these eight diagrams 
(including their complex conjugates), see Appendix \ref{app.1}.
These parameters depend on the TLL parameter $g$ and
on the dimensionless temperature $T/\varepsilon$.

\begin{figure}
\begin{center}
\epsfig{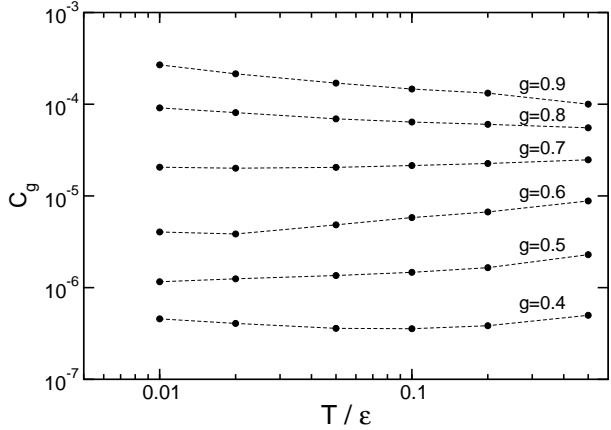}
\caption{Temperature dependence of the dimensionless 
parameter $C_g$ in 
Eq.~(\ref{fingam}) for various $g$ and $D=10\varepsilon$. Dashed lines are 
guides to the eye only. Notice the double-logarithmic scales.
\label{fig.acg} }
 \end{center}
\end{figure}

Of primary interest is then the temperature dependence of the 
self-consistently determined linewidth parameter $\gamma$.  
Unfortunately, as we discussed above, 
it seems impossible to evaluate $A_g,B_g,$ and
$C_g$ analytically, even in the non-interacting limit $g=1$.
However, numerically we can obtain them
for given $(g,T)$, see Appendix \ref{app.1} for details, and we
find $B_g\approx A_g < C_g$. Since the 
master equation approach holds only for $\gamma\ll \varepsilon$,
it is clear that the $B_g$ term in Eq.~(\ref{fingam}) can be neglected
for all practical purposes.  Numerical results  
for $A_g$ and $C_g$ for different $g$ and $T$ are shown in 
Table~\ref{tab.acg} and Fig.~\ref{fig.acg}, respectively. 
Equation (\ref{fingam}) indicates that $C_g$ follows from the large-$\gamma$
behavior of $\Gamma_{\lambda}^{\nu,(2)}$, while  
determining $A_g$ requires the small-$\gamma$ solution of the relevant
triple integral. Unfortunately, the latter is numerically rather expensive at
low temperatures, and hence we can specify $A_g$ only for moderately
low $T$, see Table~\ref{tab.acg},
while Fig.~\ref{fig.acg} covers our results for $C_g$  
down to $T=0.01 \varepsilon$. 
Evidently, the temperature dependence of $C_g$ becomes
rather weak at low temperatures,
$C_g(T) \simeq {\rm const.}$, an observation supported
by analytical arguments given in Appendix \ref{app.1}.
Based on these arguments, we expect that even for $g=0.9$, where
Fig.~\ref{fig.acg} suggests a significant $T$-dependence, at sufficiently
low $T$ the quantity $C_g$ becomes independent of temperature.

\begin{table*}
\begin{ruledtabular}
\begin{tabular}{lccc}
& $T=0.1 \varepsilon$ & $T=0.2\varepsilon$ & $T=0.5\varepsilon$ \\
\hline
$g=0.4$ & $3.37 \times 10^{-8}$ & $-1.56 \times 10^{-8}$ & $-1.12 \times 10^{-8}$ \\
\hline
$g=0.5$ & $3.00 \times 10^{-8}$ & $-2.02 \times 10^{-8}$ & $-6.07 \times 10^{-8}$ \\
\hline
$g=0.6$ & $2.35 \times 10^{-8}$ & $9.71 \times 10^{-9}$ & $2.83 \times 10^{-7}$ \\
\hline
$g=0.7$ & $-1.97 \times 10^{-8}$ & $2.32 \times 10^{-8}$ & $1.34 \times 10^{-6}$ \\
\hline
$g=0.8$ & $-1.42 \times 10^{-7}$ & $7.68 \times 10^{-8}$ & $3.86 \times 10^{-6}$ \\
\hline
$g=0.9$ & $-3.76 \times 10^{-7}$ & $7.35 \times 10^{-7}$ & $7.85 \times 10^{-6}$ \\
\end{tabular}
\end{ruledtabular}
\caption{Numerical results for the dimensionless 
parameter $A_g$ in Eq.~(\ref{fingam}) 
for various $g$ and $T$ at $D=10\varepsilon$. \label{tab.acg}}
\end{table*}

We mention in passing that in contrast to the 
first-order contributions in Sec.~\ref{sec31}, 
the above indirect CST contributions
do not obey detailed balance. 
 This is true although the correlation functions
entering these rates in Table \ref{tab.diag} still have
the reflection property.
The violation of the detailed balance relation
follows directly by inspection of the triple integrals in
Table \ref{tab.diag}, i.e., forward and backward 
rates are not linked by a relation of the form (\ref{detbalance}).
Of course, the total rates and populations of the states are
still linked by detailed balance, but there is no reason
why individual rates should obey Eq.\  (\ref{detbalance}). 
A simple example for this fact is given by superexchange
rates in electron transfer theory,\cite{Weiss99,hu} see also 
Ref.~\onlinecite{Fubini04} for related observations. 

\subsection{Direct CST rates} \label{subsec.direct}

The direct CST rates  $\Gamma_{CST}^{\nu}(n)$ for a 
transition $N\rightarrow N-2$
have only contributions of at least order $\Gamma_0^2$. There are two possible
 transition processes CST1 and CST2, see Table \ref{tab.paths}
and Figs.~\ref{fig.plan} and \ref{fig.cst}(b). 
 The corresponding transition rates are simply 
 \begin{equation}\label{cst}
 \Gamma_{CST,1}^{f}  (n) = - \Gamma_{R,7}^{f,(2)}(n) ,\quad
\Gamma_{CST,2}^{f}  (n) = - \Gamma_{R,8}^{f,(2)}(n)  .
 \end{equation}
These relations hold not only on resonance but in general. 
 Since CST transitions do not alter the stationary population
 $p(n)$, they do not enter the linewidth (\ref{linewidth}). 
 The rates $\Gamma_{CST}^{\nu}(n)$ {\sl per se\/} 
 are also not subject to a detailed balance relation (\ref{detbalance}). 
  Note that the cotunneling diagram in Fig.~\ref{fig.ust}(b) 
is subleading on resonance and therefore not taken into account here.

\subsection{Linewidth}

On resonance, $E_{\lambda \nu}(n)=0$, and hence
  forward and backward rates are equal, 
\begin{equation}
\Gamma_{\lambda}^{f,(1)/(2)}=\Gamma_{\lambda}^{b,(1)/(2)}\equiv
\Gamma^{(1)/(2)} . 
\end{equation}
Neglecting the $B_g$-term as explained above,
the self-consistency equation for $\gamma$ can then be written as 
\begin{equation}
\gamma = 4 (\Gamma^{(1)} + \Gamma^{(2)}) = 4 \Gamma^{(1)} +\frac{4 \Delta^4}
{\varepsilon^3}
\left(  -\frac{\varepsilon}{\gamma} A_g + C_g\right) .
\end{equation}
This quadratic equation has two real solutions, as long as the dimensionless
parameter $\xi<1$, where
\begin{equation}
\xi= \frac{\Delta^4 \varepsilon^4 A_g}
{(\varepsilon^3\Gamma^{(1)} + \Delta^4 C_{g})^2}  .
\label{xi}
\end{equation}
If indeed $\xi<1$, these solutions are given by
\begin{equation}\label{gammasolution}
\gamma_{\pm} = 4 \left[ \Gamma^{(1)} + \frac{\Delta^4 C_{g}}{\varepsilon^3} \right] 
\frac{1 \pm \sqrt{1-\xi}}{2} .
\end{equation}
For $\xi \ll 1$, we thus have
\begin{equation} \label{lingam}
\gamma \equiv \gamma_+ \approx 4\left(\Gamma^{(1)} +
 \frac{\Delta^4 C_{g}}{\varepsilon^3}\right),
\end{equation}
and $\gamma_-=\xi \gamma_+/4$.
For $\xi > 1$, the two solutions are complex valued, and $\gamma$ acquires an
imaginary part. If this happens, the description in terms
of the master equation in combination with the Weisskopf-Wigner
approximation is questionable and not used below.
We identify the linewidth with the solution $\gamma_+$, since 
we recover the well-known result\cite{Furusaki}
$ \gamma = 4 \Gamma^{(1)}$ when second-order rates are neglected, 
 while then $\gamma_-=0$.
The requirement $\xi<1$  typically results in a 
temperature $T_l$ determined by 
\[
 \Delta^4 \frac{A_g(T_l)}{\varepsilon^2} = \left( \Gamma^{(1)}(T_l) + 
 \frac{\Delta^4 C_g}{\varepsilon^3} \right)^2 ,
\]
 below which our Weisskopf-Wigner theory becomes problematic.
This equation yields $T_l$  provided $A_g(T)$ and $C_g$ 
(assumed independent of temperature) 
are known.  However, since no reliable low-$T$
estimates for $A_g$ are available, 
see Table \ref{tab.acg}, it is often difficult to provide 
good estimates for $T_l$.  Of course, the validity of the master
equation in addition always  requires $G_{\rm max} \ll G_0$. 
In what follows, the parameter $C_g$ is assumed to be temperature independent 
and given by its value at $T/\varepsilon=0.01$ in Fig.~\ref{fig.acg}.
Although we expect $C_g(T)$ to be constant (see above),
 we cannot rule out that this is approximative.

\begin{figure}
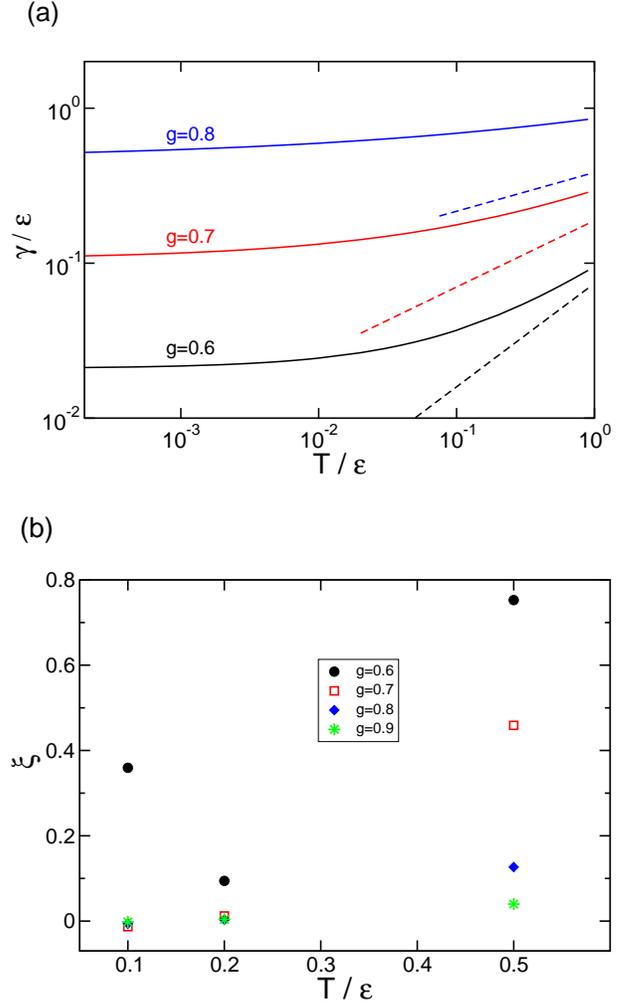

\begin{center}
\epsfig{figure=f5a.eps,width=80mm,angle=0,keepaspectratio=true}
\\[5mm] 
\epsfig{figure=f5b.eps,width=80mm,angle=0,keepaspectratio=true}
\caption{(Color online) (a) Linewidth $\gamma$ and (b) 
the parameter $\xi$ in Eq.~(\ref{xi}) as a function of
$T$ for different $g$ at $\Delta=6 \varepsilon, D=10 \varepsilon$.  
Dashed lines in (a) represent  $4 \Gamma^{(1)} \propto T^{-1+1/g}$. 
Notice the double-logarithmic scales. (Using these parameters,
the master equation approach breaks down for $g=0.9$.)
 \label{fig.linewdel60}}
 \end{center}
\end{figure}

\begin{figure}
\begin{center}
\epsfig{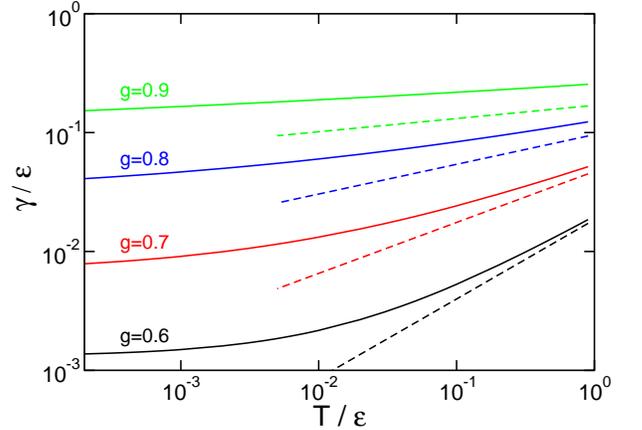} 
\caption{(Color online) Same as Fig.~\ref{fig.linewdel60}(a), but for
 $\Delta=3 \varepsilon$.  
 \label{fig.linewdel30}}
 \end{center}
\end{figure}

\begin{figure}
\begin{center}
\epsfig{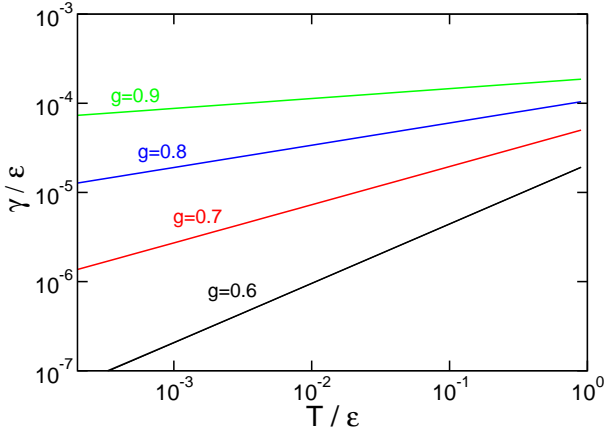}
\caption{(Color online) Same as Fig.~\ref{fig.linewdel30}, 
but for $\Delta=0.1 \varepsilon$.  Dashed lines
describing $4\Gamma^{(1)}$ coincide with the full result 
for $\gamma$.  \label{fig.linewdel01}}
 \end{center}
\end{figure}

The linewidth $\gamma(T)$ now consists of two contributions. 
The first term in Eq.~(\ref{lingam}) is $\propto T^{1/g-1}$, while the second
term is  constant due to the $T$ independence of $C_g$.
 This implies a crossover from power-law scaling of $\gamma(T)$ to a
basically constant $\gamma$ as the temperature is lowered.
This crossover depends in an essential way on the tunneling amplitude $\Delta$.
Results for $\gamma(T)$ from Eq.~(\ref{lingam}) 
are shown in Fig.~\ref{fig.linewdel60}(a),
taking  $\Delta =6\varepsilon$ and $D=10\varepsilon$.
Figure \ref{fig.linewdel60}(b) shows (for several $T$) 
that the validity condition $\xi<1$, with $\xi$ defined in Eq.~(\ref{xi}),
is indeed fulfilled.
This choice for $\Delta$ reflects 
rather transparent barriers, where CST effects are clearly pronounced.
Obviously, at low $T$, Fig.~\ref{fig.linewdel60} suggests that  
$\gamma$ is essentially independent of temperature. 
This behavior is most pronounced for strong interactions
(small $g$).  Going towards less transparent barriers,
$\gamma(T)$ is shown in Fig.~\ref{fig.linewdel30} for $\Delta=3\varepsilon$.
In Sec.~\ref{sec5}, we argue that the hopping amplitudes $\Delta$  appropriate
for the experiment in Ref.~\onlinecite{Postma01} and for the
 numerical simulations in Ref.~\onlinecite{Hugle04} are
comparable to this value.  Now a
crossover from the UST power law at high temperatures to 
an approximately $T$-independent behavior at low $T$
becomes apparent.  For  $g \le 1/2$, we now find $\xi > 1$,
implying that our approach breaks down for such interactions. 
Finally,  for very high barriers, $\Delta\to 0$,  
the linewidth is {\sl always\/}
dominated by the UST term, in accordance with standard
reasoning.\cite{Furusaki98}  This is
illustrated in Fig.~\ref{fig.linewdel01}, where results for $\gamma$
at $\Delta = 0.1 \varepsilon$ are depicted.
In this regime, higher-order corrections  are obviously negligible. 

Remarkably, for  weak interactions, 
the linewidth $\gamma$ is always dominated by the UST result 
over the {\sl entire} range of temperatures where
the master equation approach is valid ($\xi < 1$).
CST effects then apparently do not have a finite domain of 
observability in the limit of weak interactions.  We can therefore
rationalize that previous calculations that essentially expand around 
the noninteracting case \cite{Nazarov03,Polyakov03,meden} do not 
observe a clear CST power-law scaling.  The fate of CST effects
near the noninteracting limit $g=1$ will be further discussed at the end of
Sec.~\ref{sec4}.  We conclude that {\sl for CST effects to be observable,
it is essential to allow for rather transparent barriers, 
a finite level spacing, and intermediate-to-strong interactions.}  
The parameter regime where CST plays a role is
therefore rather narrow. 

\section{Temperature
 dependence of  conductance peak} \label{sec4} 

The linear conductance $G$ follows 
directly from Eqs.~(\ref{currentx1}) and (\ref{currentx2}) 
by performing the derivative with respect to the transport voltage $V$. 
Moreover, since we are interested in the 
conductance maximum at resonance, we can put $\mu(n)=\mu(n+1)=0$. 
For the first-order rates, the 
detailed balance relation (\ref{detbalance}) can be exploited.  
For the second-order rates, we find the relation 
$d \Gamma^{f,(2)}/dV  = - d \Gamma^{b,(2)}/dV$. Finally  
setting $V=0$, we obtain
\begin{equation}
G_{\rm max}(T) = G_{{\rm max},A}+G_{{\rm max},B}  ,
\end{equation}
where
\begin{eqnarray}
G_{{\rm max},A}  &=&  \frac{e^2}{T \gamma} \left(\Gamma^{(1)}
\right)^2 \, , \nonumber \\
G_{{\rm max},B}  & = &
-e\Gamma^{(2) \prime} - 2 e \Gamma_{CST}^{\prime} + \frac{e^2}{T\gamma}
\Gamma^{(1)}\Gamma^{(2)}.
\end{eqnarray}
Here the prime denotes the derivative $d/dV$, taken at $V=0$,
 and we omit the arguments of the
rates, since on resonance they are all equal.
The dominant term is the first term in  $G_{{\rm max},A}$.
In order to show that the second term $G_{{\rm max},B}$ is always negligible 
against $G_{{\rm max},A}$, 
it is instructive to evaluate 
 the second class of diagrams ($k=3,4,5,6$) in Table
  \ref{tab.diag} within a simple approximation.   
  Since at long times the Ohmic part in
 $W_{\Sigma}(t)$ cancels out in the square-bracketed terms 
of the corresponding
rate expressions in Table \ref{tab.diag}, one may replace $W_\Sigma(t)$ by 
$W_{\rm  dot}(t)$ in those square brackets. After this replacement, the Fourier
expansion method discussed in the Appendix applies to all eight diagrams, 
 with Fourier sums including terms like $dK_{\rm R/L}(E)/dE$.
These sums can easily be calculated numerically. 
For all cases that are described below (and many more not shown here), 
we find that 
$G_{{\rm max},B}$ is numerically zero.  
Although it looks like the direct CST rates $\Gamma_{CST}$
have effectively no influence in the end, this is not true since
they exactly cancel certain contributions to the conductance coming from the 
indirect rate $\Gamma^{(2)}$.  In that sense, the direct CST diagram, jumping by
two steps along the diagonal of the reduced density matrix
via an intermediate diagonal state, see Fig.~\ref{fig.plan}, but
without cutting the diagram in the intermediate state, is crucial
for CST effects in the conductance maximum $G_{\rm max}$.
Such diagrams are known to cause important effects in other
systems,\cite{hu} but were previously not taken into account since the main
focus was on the limit $\Delta\to 0$. 

\begin{figure}
\begin{center}
\epsfig{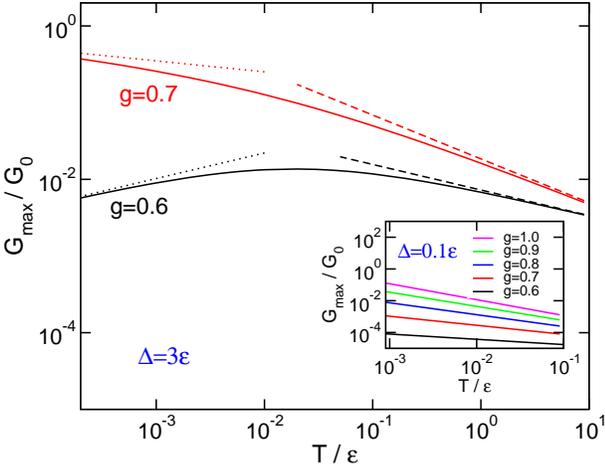}
\caption{(Color online) Temperature dependence of the conductance maximum
 $G_{\rm max}$ (solid lines) for $\Delta=3
\varepsilon$ and TLL parameters $g=0.6$  and $g=0.7$.  
Note the double-logarithmic scales.  Dotted (dashed) lines represent
the CST (UST) power law $G_{\rm max} \propto T^{\alpha_{\rm CST}}$ 
($G_{\rm max} \propto T^{\alpha_{\rm UST}}$). 
Inset: $G_{\rm max}$ for $\Delta=0.1 \varepsilon$ 
 and $g$ between $g=0.6$ (bottom) and $g=1.0$ (top). The slopes
coincide with $\alpha_{\rm UST}=-2+1/g$.
 \label{fig.conddel30}}
 \end{center}
\end{figure}

Hence we find for the conductance maximum  
\begin{equation} \label{condmax}
G_{{\rm max}}(T) \simeq \frac{e^2}{T \gamma} \left(\Gamma^{(1)}
\right)^2 .
\end{equation}
Judging from our numerical results for $G_{\rm max,B}$, 
the '$\simeq$' should in fact be replaced by
an exact equality, although we have no analytical proof for
this statement. 
For $T>T_c$, we find $\gamma\propto T^{1/g-1}$, 
leading to the UST result (\ref{seq}).
However, for $T\alt T_c$,  $\gamma(T)$ stays approximately constant, 
and hence an approximate power-law behavior follows,
\begin{equation}            \label{gcst}
G_{{\rm max}} \propto T^{2/g-3},
\end{equation}
with the CST exponent $\alpha_{\rm CST}=-3+2/g$. 
We stress that Eq.~(\ref{gcst}) is not meant 
in the sense of universal power-law scaling behavior.
The crossover between UST and CST-dominated regimes
occurs around a temperature $T_c$ discussed below.
 The effective doubling in the exponent reflects the physics of this
correlated transport process. For $T \ll \varepsilon$, plasmon modes excited 
on the island will correlate electrons in both leads. Since each lead
has a end-tunneling density of states $\propto E^{1/g-1}$, 
correlated transport leads to an effective doubling in the exponent due to the
presence of two leads.\cite{Thorwart02}
Results for $G_{{\rm max}}(T)$ at $\Delta=3 \varepsilon$ are shown in 
Fig.~\ref{fig.conddel30} and follow Eq.~(\ref{gcst}) at low $T$.
For $g\agt 0.8$, the master equation breaks down ($G_{\rm max}$ exceeds
$G_0$), while for $g\leq 1/2$, the validity condition $\xi<1$ is violated. 
Nevertheless, there is a well-defined region of applicability, where 
CST effects are important and observable.
Finally, for $\Delta=0.1 \varepsilon$, the expected 
UST scaling is recovered, see the inset of Fig.~\ref{fig.conddel30}. 

\begin{figure}
\begin{center}
\epsfig{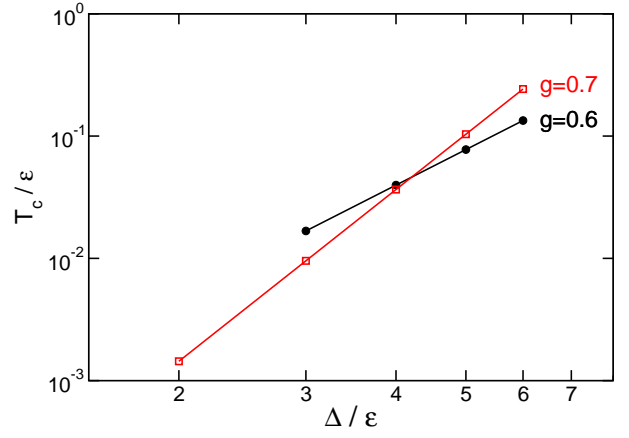}
\caption{Crossover temperature $T_c$  
separating the UST ($T>T_c$) and CST ($T<T_c$) 
regimes ($D=10 \varepsilon$) for different $\Delta$ and $g=0.6, 0.7$.
 \label{fig.tcg}}
 \end{center}
\end{figure}

The crossover between the UST and CST regimes is
 characterized by a temperature $T_c=T_c(\Delta,g)$, which in turn  follows
from the condition that both contributions to $\gamma$
in Eq.~(\ref{lingam}) be equal,  $\Gamma^{(1)}(T_c) = \Delta^4 C_{g}/
\varepsilon^3$. 
Although $\Gamma^{(1)}$ contains the sum over all Fourier modes, 
see Eq.~(\ref{ustrate}), the zero mode $p=0$ dominates 
 on resonance, implying the condition 
$d_0 \Gamma_{{\rm Ohm}}(0)= \Delta^4 C_{g}/\varepsilon^3$.
This  leads to the crossover temperature
\begin{equation}\label{tcc}
T_c=\left( \frac{\Delta^2 C_{g}}{\eta_g\varepsilon^3}\right)^{g/(1-g)}  ,
\end{equation}
where we use the abbreviation
\[
\eta_g = \frac{(1-e^{-\varepsilon/D})^{1/g} \Gamma^2(1/2g)}{4 D
\Gamma(1/g)} (2 \pi  /D)^{1/g-1} .
\]
In Fig.~\ref{fig.tcg}, results for the crossover temperature $T_c$
are shown as a function of $\Delta/\varepsilon$ 
for the TLL parameters $g=0.6$ and $g=0.7$, 
always respecting the validity conditions
$\xi(T_c)<1$ and $G_{\rm max} (T_c) \ll G_0$. 
Generally, $T_c$ increases when increasing $\Delta$ and/or
decreasing $g$, i.e., for more transparent barriers and/or
stronger interactions. Apparently, for weak interactions, $g\alt 1$,
 the $\Delta$ dependence of $T_c$ becomes very steep, restricting
the CST regime to extremely low temperatures for reasonable $\Delta$. 
In practice, at such low temperatures coherent resonant tunneling processes
dominate, rendering CST effects unobservable.  For stronger interactions,
however, CST effects can be pronounced even for moderately 
transparent barriers at low temperatures.

\section{Discussion} \label{sec5}

By using a master equation approach, linear
transport in a TLL with two tunneling barriers forming a
quantum dot has been studied. We find an approximate power-law
 temperature dependence of the peak conductance in the linear transport regime,
with a characteristic $g$-dependent exponent, where $g$ is the TLL 
parameter.
By including second-order contributions to the tunneling rates in 
combination with a self-consistent Weisskopf-Wigner regularization,
a comprehensive picture has been obtained. 
For temperatures below the plasmon level spacing 
$\varepsilon=\pi v_F/(g x_0)$ for dot size $x_0$, the 
dominant transport mechanism depends
 on the transparency $\Delta$ of the barriers.
For sufficiently transparent barriers, $\Delta \agt \varepsilon$, we find
that correlated sequential tunneling (CST) is important, leading
to an approximate power law with exponent $\alpha_{\rm CST}=-3+2/g$,
while for very high barriers ($\Delta \to 0$),
the uncorrelated sequential tunneling (UST)
exponent $\alpha_{\rm UST}=-2+1/g$ is recovered.  
Note that despite the large $\Delta$ necessary to reach the CST regime,
the master equation should still apply as the peak conductance remains
small.  We have determined the crossover temperature
$T_c$ separating both regimes. As a function of the
physical parameters ($g,T,\Delta,\varepsilon,D)$, 
it is given by Eq.~(\ref{tcc}), where the dimensionless number
 $C_g$ can be  read off from Fig.~\ref{fig.acg}.
For $T<T_c$, the CST mechanism is effective, while for
$T>T_c$, the UST picture is recovered.
The peak conductance in the CST regime has been derived and is 
given by Eq.~(\ref{condmax}), with the well-known first-order rates
$\Gamma^{(1)}$ specified in Eq.~(\ref{ustrate}). The linewidth
parameter $\gamma$ as a function of the physical
parameters is given in Eq.~(\ref{lingam}).  These relations
allow for a comparison to existing work and provide
estimates for the parameter regime where such effects should be
of importance.
Notice that the modification of the standard picture of sequential
tunneling here arises due to a renormalization
of $\gamma$ by higher-order processes, and we have
given a physical explanation for this mechanism above.

Our findings regarding approximate CST power laws in $G_{\rm max}(T)$ are
consistent with recent numerically exact real-time quantum Monte Carlo
simulations \cite{Hugle04} and also with experimental observations. 
Let us first discuss the  
experimental work on SWNTs,\cite{Postma01} where the 
conductance through the dot followed the CST power law
(\ref{gcst}). From the exponent, the interaction strength was deduced to be
$g=0.54$, corresponding to a TLL parameter $g_{\rho}=0.23$ for
the charged sector of the effective four-channel
 TLL theory of SWNTs.\cite{egger,kane} 
The quantum dot was formed by two nearby buckles acting as 
tunneling barriers. Since CST effects are predicted only
for quite transparent barriers, it is instructive to estimate  $\Delta$.
This is  simpler for a single buckle used in
earlier experiments,\cite{yao} where also a power-law linear conductance
$G_{\rm 1B}(T)$ was observed.
The single-barrier case is analytically solvable,\cite{Kane92} and for 
a high barrier,  
\[
\frac{G_{\rm 1B}}{4 G_0} =
\frac{\pi^{5/2} \Gamma (1+1/g)}{2 \Gamma (1/2+1/g)}
(\Delta/D)^2
(\pi T/D)^{2/g-2} .
\]
The measurements\cite{yao} for $G_{\rm 1B}$ yield together with $g=0.54$
a barrier transparency $\Delta \approx 45$~meV, taking 
a bandwidth of $D=0.5$~eV; see Ref.~\onlinecite{dekker}.
Assuming that the buckles have similar features when two
are designed in a row, we can now establish a connection
to  the double-barrier case. In Ref.~\onlinecite{Postma01}, 
 $\varepsilon = 38$~meV was measured, 
yielding
$\Delta \approx 1.2 \varepsilon$, consistent with our conclusions
above.  For CST to be operative, one needs finite level spacing, 
low (but not too low) temperatures, 
and not too high barriers.  These conditions apparently
were met in the SWNT experiments in Ref.~\onlinecite{Postma01}.  
Let us then comment on the Monte Carlo results of Ref.~\onlinecite{Hugle04}, 
where also the CST power law (\ref{gcst}) has been found. 
For a direct comparison, we determine $\Delta$
for the potential strength $U_{\rm imp}$  used in Fig.~4 of
Ref.~\onlinecite{Hugle04}.
Using Eq.~(\ref{gamma0}), for the simulation parameters 
$g=0.6, U_{\rm imp}=0.2 D$,  and $E_s=\pi D/20$,
we find $\Delta \simeq 3.3 \varepsilon$, again consistent with
our conclusions.
We note in passing that Eq.~(\ref{gcst}) has been obtained 
in Ref.~\onlinecite{Thorwart02} starting from finite-range interactions among
the electrons. The divergence has been regularized by summing up a selection 
of  higher-order terms.  However, this selection was
too strict,\cite{remark4} leading to the incorrect conclusion 
that finite-range interactions would be  
a prerequisite for CST to occur. As shown here, also zero-range
interactions  suffice, as long as the tunneling barriers are not too 
high. 

We emphasize that for Fermi liquid leads ($g=1$) and $T\ll \varepsilon$,
one finds $G_{\rm max}\propto T^{-1}$ both within a UST and a CST analysis.
Several researchers recently approached the double-barrier TLL problem by
considering weak interactions, $g$ close to 1, without evidence for CST 
scaling.\cite{Nazarov03,Polyakov03,meden}
As we have discussed in Sec.~\ref{sec4}, in this weak-interaction
limit, the crossover temperature $T_c$ very quickly goes to zero when
decreasing $\Delta$, excluding CST effects for weak interactions
even for finite level spacing. Put differently, for $g$ close to 1,
the master equation approach for large $\Delta$ 
will always break down ($\xi$ becomes larger than 1) upon lowering $T$
{\sl before\/} CST sets in. Therefore our results are in fact consistent
with previous results.\cite{Nazarov03,Polyakov03,meden}
Furthermore, the functional renormalization group calculation of
Ref.~\cite{meden} reported traces of an ``apparent''
power law (as opposed to true scaling) that could be
linked to the CST mechanism.
Finally, Komnik and Gogolin\cite{Komnik03} presented an exact solution
of a related model at the point $g=1/2$. 
In their model, however, there is {\sl no
sequential tunneling regime at all}, and therefore
we believe that this represents a non generic situation
that has nothing to say about the issues at stake here. 
This point has also been clarified in other recent 
publications.\cite{Hugle04,meden}
Unfortunately, this also excludes the possibility of an independent
analytical check of our results.

To conclude, we hope that these novel features of interacting
one-dimensional electrons will stimulate other theoretical
work as well as further experimental
checks of the CST versus UST picture put forward here.

\begin{acknowledgments}
We thank  A.O.~Gogolin,
V.~May, Yu.~Nazarov, U.~Weiss, and especially J.~Stockburger for discussions. 
This work has been supported by the EU network DIENOW, the SFB TR/12,
and the Gerhard-Hess program of the DFG. 
\end{acknowledgments}

\appendix

\section{Fourier expansion scheme \label{app.1}}

\begin{figure}
\begin{center}
\epsfig{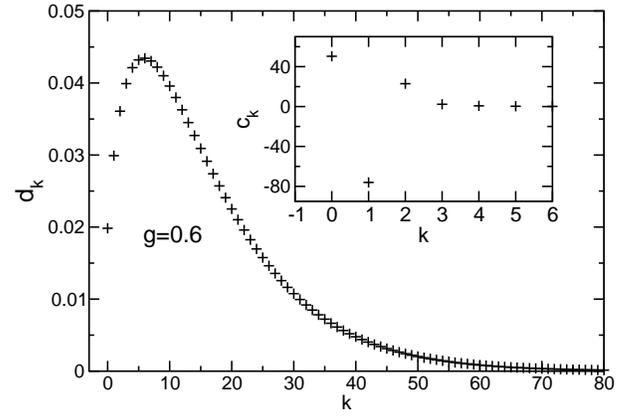}
\caption{Fourier components $d_k$ and $c_k$ (inset) 
for $g=0.6$ and $T\ll \varepsilon$. 
 \label{fig.fourierdc}}
 \end{center}
\end{figure}

\begin{figure}
\begin{center}
\epsfig{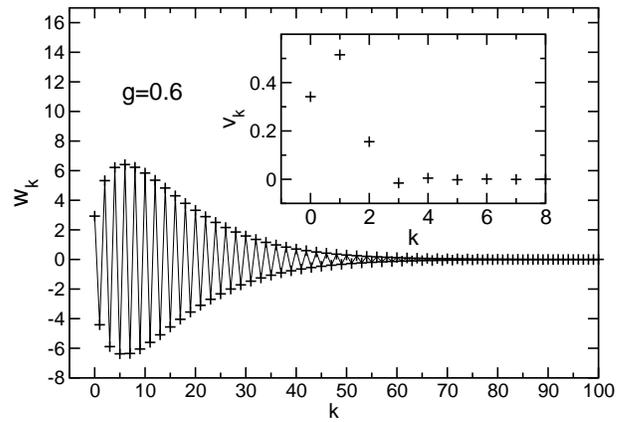}
\caption{Same as Fig.~\ref{fig.fourierdc}
but for the Fourier components $w_k$ and $v_k$ (inset).
 \label{fig.fouriervw}}
 \end{center}
\end{figure}

To evaluate the rate expressions in Sec.~\ref{sec3},
 we often need the quantities  $e^{\pm W_{\rm dot}(t)}$
and $e^{\pm W_\Delta(t)}$.
 Exploiting the periodicity of these correlation functions with period
$\tau_\varepsilon=2\pi/\varepsilon$,
 it is convenient to perform a Fourier expansion,
\begin{eqnarray*}
e^{W_{\rm dot}(t)}&=&\sum_{p=-\infty}^{\infty}c_p(\varepsilon)
e^{-ip\varepsilon t},\\
e^{-W_{\rm dot}(t)}&=&\sum_{p=-\infty}^{\infty}d_p(\varepsilon)
e^{-ip\varepsilon t},\\
e^{-W_{\Delta}(t)}&=&\sum_{p=-\infty}^{\infty}v_p(\varepsilon)
e^{-ip\varepsilon t},\\
e^{W_{\Delta}(t)}&=&\sum_{p=-\infty}^{\infty}w_p(\varepsilon)
e^{-ip\varepsilon t}.
\end{eqnarray*}
We note that Eqs.~(\ref{Wdot}) and (\ref{Wdelta})  imply that
these Fourier coefficients are real. For low temperatures $T$,
but keeping leading corrections in $y=e^{-\varepsilon/T}$,
one finds, with the Heaviside function $\theta(x)$:
\begin{eqnarray*}
c_{k}(\varepsilon,T)&=&c_{k}- \frac{1}{g}y
\left(c_{k-1}+c_{k+1}+2 c_{k} \right)+ {\mathcal O}(y^2)\;, \\
 c_{k}&=&\theta(k) (-1)^{k}\frac{(1-e^{-\varepsilon/D})^{-1/g}}
 {k!}  \nonumber \\
 & & \times  \frac{\Gamma(1/g+1)}{\Gamma(1/g-k+1)}e^{-k\varepsilon/D}
, \\ d_{k}(\varepsilon,T)& = &d_{k}+ \frac{1}{g}y
\left(d_{k-1}+d_{k+1}-2 d_{k} \right)+ {\mathcal O}(y^2),
 \\ d_{k}&=&\theta(k) \frac{(1-e^{-\varepsilon/D})^{1/g}} {k!}
\frac{\Gamma(1/g+k)}{\Gamma(1/g)}e^{-k\varepsilon/D}  , \\
v_{k}(\varepsilon, T)& = & (-1)^k \chi c_k(\varepsilon,T)  , \\
 \nonumber \chi & = &  \frac{(1+e^{-\varepsilon/D})^{-1/g}}
{(1-e^{-\varepsilon/D})^{-1/g}}\, , \\
w_{k}(\varepsilon, T)& = & (-1)^k \chi^{-1} d_k(\varepsilon,T) .
\end{eqnarray*}
Note that $c_0 d_0=v_0 w_0 =1$. 
The Fourier coefficients are shown in Figs.\  \ref{fig.fourierdc}
 and \ref{fig.fouriervw} for $g=0.6$ and  $T \ll \varepsilon$.
In that case, the $T=0$ form of the correlation functions (\ref{Wdot}) and
(\ref{Wdelta}) can be taken.  Since the number of non zero Fourier
coefficients is not exceedingly large, a quick, very accurate,
 and reliable numerical
scheme can be implemented for the evaluation of diagrams of the first class. 

Next, as mentioned in Sec.~\ref{sec32}, we show in detail
how diagrams of the first class ($k=1,2,7,8$) in Table \ref{tab.diag}
have been handled,
taking  $\Gamma_{R,7}^{f,(2)}(n)$ as concrete example. 
We consider the expressions on resonance and for low temperature
$T\ll \varepsilon$.  After inserting the above Fourier expansions,
the integration over $\tau_2$ can be performed directly.
Since $w_0 v_0=1$, 
the `$-1$' in the square bracket expression in
 $\Gamma_{R,7}^{f,(2)}$, see Table \ref{tab.diag},
is exactly canceled by the corresponding Fourier term. We find 
\begin{widetext}
\begin{eqnarray}
\Gamma_{R,7}^{f,(2)}(n) & = &  -2 \frac{\Delta^4}{16}  \mbox{\rm Re }
\sum_{k,l=0}^{\infty} \,\, \sideset{}{^\prime}\sum_{m,r,p,q=0}^{\infty}
\frac{d_k d_l w_m v_r v_p w_q}{\gamma - i \varepsilon_{-mr-pq} }
\int_0^{\infty} d\tau_1 e^{-W_{\rm Ohm}^*
(\tau_1)+i\varepsilon_{-k-mp} \tau_1} 
\int_0^{\infty} d\tau_3 
e^{-W_{\rm Ohm}(\tau_3)+i\varepsilon_{-l-pq} \tau_3}\, ,  \nonumber \\ 
\end{eqnarray}
\end{widetext}
where we have introduced the notation 
$\varepsilon_{-mr-pq}= (-m+r-p+q) \varepsilon$, and
analogously for $\varepsilon_{-k-mp}$. 
The prime in $\sideset{}{^\prime}\sum$ 
indicates that the term with $m=r=p=q=0$ is excluded
from the sum.
Next, we define the dimensionless kernels 
[cf.~Eqs.~(\ref{ohmic}) and (\ref{ohmicrate})]
\begin{eqnarray}
K_{\rm R} (E) & = & \varepsilon 
 \int_0^{\infty} d\tau 
e^{-S_{\rm Ohm}(\tau)} \cos \left[ E \tau -
 R_{\rm Ohm}(\tau)\right] \nonumber \\
 & = & \frac{2 \varepsilon}{\Delta^2} \Gamma_{\rm Ohm} (E) ,  \\
K_{\rm I} (E) & = & \varepsilon \int_0^{\infty} d\tau e^{-S_{\rm Ohm}(\tau)} 
\sin \left[ E \tau - R_{\rm Ohm}(\tau)\right] , \nonumber 
\label{kernk}
\end{eqnarray}
such that
\begin{widetext}
\begin{eqnarray*}
\Gamma_{R,7}^{f,(2)}(n) & = &  - 2 \frac{\Delta^4}{16 \varepsilon^2}
\sum_{k,l=0}^{\infty} \, \, \sideset{}{^\prime}\sum_{m,r,p,q=0}^{\infty}
\frac{d_k d_l w_m v_r v_p w_q}{\gamma^2 + \varepsilon_{-mr-pq}^2 }
\left\{
\gamma K_{\rm R}(-\varepsilon_{kmp})K_{\rm R}(\varepsilon_{-l-pq}) 
 - \gamma K_{\rm I}(-\varepsilon_{kmp})
K_{\rm I}(\varepsilon_{-l-pq})  \right. \\
& & \left.
- \varepsilon_{-mr-pq} \left[  K_{\rm R}(-\varepsilon_{kmp})K_{\rm
I}(\varepsilon_{-l-pq})
+ K_{\rm I}(-\varepsilon_{kmp})K_{\rm R}(\varepsilon_{-l-pq}) \right]
\right\}  .
\end{eqnarray*}
\end{widetext}
Since one needs to self-consistently enforce
 $\gamma \ll \varepsilon$, we can simplify this
expression to the form
\begin{equation} \label{funcdep}
\Gamma_{R,7}^{f,(2)}(n) = \frac{\Delta^4}{\varepsilon^3}
 \left(- \frac{\varepsilon}{\gamma} A_{g,7} +
\frac{\gamma}{\varepsilon} B_{g,7}
 + C_{g,7} \right)   ,
\end{equation}
with the $\gamma$-independent terms
\begin{widetext}
\begin{eqnarray}
A_{g,7} & = & \frac{1}{8} \sum_{k,l=0}^{\infty} \, \,
 \sideset{}{^\prime}\sum_{r,p,q=0}^{\infty}
d_k d_l d_{r-p+q} c_r c_p d_q
\left\{
K_{\rm R}(-\varepsilon_{krq})K_{\rm R}(\varepsilon_{-l-pq}) -
K_{\rm I}(-\varepsilon_{krq})K_{\rm I}(\varepsilon_{-l-pq})  \right\} \nonumber  \\
B_{g,7} & = & - \frac{1}{8} \sum_{k,l=0}^{\infty} \, \, \, 
\sideset{}{^{\prime \prime}} \sum_{m,r,p,q=0}^{\infty}
\frac{d_k d_l d_{m} c_r c_p d_q}{(-m+r-p+q)^2}
(-1)^{m+r+p+q}
\left\{
K_{\rm R}(-\varepsilon_{kmp})K_{\rm R}(\varepsilon_{-l-pq}) \right. \nonumber \\
& & \left.  
- K_{\rm I}(-\varepsilon_{kmp})K_{\rm I}(\varepsilon_{-l-pq})  \right\}  \nonumber\\
C_{g,7} & = & - \frac{1}{8} \sum_{k,l=0}^{\infty}\,  \, \, 
\sideset{}{^{\prime \prime}}
\sum_{m,r,p,q=0}^{\infty}
\frac{d_k d_l d_{m} c_r c_p d_q}{-m+r-p+q}
(-1)^{m+r+p+q} \left\{
-K_{\rm R}(-\varepsilon_{kmp})K_{\rm I}(\varepsilon_{-l-pq}) \right. \nonumber\\
& & \left. 
 -K_{\rm I}(-\varepsilon_{kmp})K_{\rm R}(\varepsilon_{-l-pq})  \right\} \, .
\label{abc1}
\end{eqnarray}
\end{widetext}
Here, the second prime in $\sideset{}{^{\prime \prime}}\sum$ denotes the sum
with the constraint $-m+r-p+q \ne 0$. 
The remaining Fourier sums are performed numerically.

For completeness, we finally summarize
the corresponding results for the remaining
diagrams of this type, $k=1,2,8$.
The final result for each diagram will be of
the form (\ref{funcdep}), with $A_{g,1}=A_{g,2}=0$ and
\begin{widetext}
\begin{eqnarray}
A_{g,8} & = & \frac{1}{8} \sum_{k,l=0}^{\infty} \, \,
\sideset{}{^\prime}\sum_{r,p,q=0}^{\infty} d_k d_l c_{r-p+q} d_r d_p c_q
\left\{ K_{\rm R}(-\varepsilon_{krq})K_{\rm R}(\varepsilon_{-lp-q}) +
K_{\rm I}(-\varepsilon_{krq})K_{\rm I}(\varepsilon_{-lp-q})  \right\},  \nonumber\\
B_{g,1} & = &  -\frac{1}{8} \sum_{k,l=0}^{\infty} \, \,\sideset{}{^{\prime}}
\sum_{m,r,p,q=0}^{\infty} 
\frac{d_k d_l d_{m} c_r c_p d_q}{(m+r+p+q)^2} (-1)^{m+r+p+q}
\left\{ K_{\rm R}(-\varepsilon_{kmp})K_{\rm R}(\varepsilon_{-lpq}) \right. \nonumber\\
& & \left. +
K_{\rm I}(-\varepsilon_{kmp})K_{\rm I}(\varepsilon_{-lpq})  \right\}  , \nonumber\\
B_{g,2} & = &  -\frac{1}{8} \sum_{k,l=0}^{\infty} \, \,\sideset{}{^{\prime}}
\sum_{m,r,p,q=0}^{\infty} \frac{d_k d_l c_{m} d_r d_p c_q}{(m+r+p+q)^2}
(-1)^{m+r+p+q}
\left\{ K_{\rm R}(-\varepsilon_{kmp})K_{\rm R}(-\varepsilon_{lpq}) \right. \nonumber\\
& & \left. -
K_{\rm I}(-\varepsilon_{kmp})K_{\rm I}(-\varepsilon_{lpq})  \right\}  , \nonumber\\
B_{g,8} & = & - \frac{1}{8} \sum_{k,l=0}^{\infty} \, \,\,\sideset{}{^{\prime
\prime}} \sum_{m,r,p,q=0}^{\infty}
\frac{d_k d_l c_{m} d_r d_p c_q}{(-m+r-p+q)^2}
(-1)^{m+r+p+q} \left\{
K_{\rm R}(-\varepsilon_{kmp})K_{\rm R}(\varepsilon_{-lp-q}) \right. \nonumber\\
& & \left. +
K_{\rm I}(-\varepsilon_{kmp})K_{\rm I}(\varepsilon_{-lp-q})  \right\} , \nonumber\\
C_{g,1} & = & - \frac{1}{8} \sum_{k,l=0}^{\infty}\, \, \sideset{}{^{\prime}}
\sum_{m,r,p,q=0}^{\infty}
\frac{d_k d_l d_{m} c_r c_p d_q}{m+r+p+q}
(-1)^{m+r+p+q} \left\{
-K_{\rm R}(-\varepsilon_{kmp})K_{\rm I}(\varepsilon_{-lpq}) \right. \nonumber\\
& & \left. -
K_{\rm I}(-\varepsilon_{kmp})K_{\rm R}(\varepsilon_{-lpq})  \right\}  , \nonumber\\
C_{g,2} & = & - \frac{1}{8} \sum_{k,l=0}^{\infty}\, \, \sideset{}{^{\prime}}
\sum_{m,r,p,q=0}^{\infty}
\frac{d_k d_l c_{m} d_r d_p c_q}{m+r+p+q}
(-1)^{m+r+p+q} \left\{
K_{\rm R}(-\varepsilon_{kmp})K_{\rm I}(-\varepsilon_{lpq}) \right. \nonumber\\
& & \left. +
K_{\rm I}(-\varepsilon_{kmp})K_{\rm R}(-\varepsilon_{lpq})  \right\}  , \nonumber\\
C_{g,8} & = & - \frac{1}{8} \sum_{k,l=0}^{\infty}\, \,
\sideset{}{^{\prime \prime}}
\sum_{m,r,p,q=0}^{\infty}
\frac{d_k d_l c_{m} d_r d_p c_q}{-m+r-p+q}
(-1)^{m+r+p+q} \left\{
K_{\rm R}(-\varepsilon_{kmp})K_{\rm I}(\varepsilon_{-lp-q}) \right. \nonumber\\
& & \left. -
K_{\rm I}(-\varepsilon_{kmp})K_{\rm R}(\varepsilon_{-lp-q})  \right\} .
\label{abc2}
\end{eqnarray}
\end{widetext}
Finally, we explain why we expect $C_g(T)$ to be $T$ independent,
 focusing on diagrams of the first class ($k=1,2,7,8)$.  
Using Eqs.~(\ref{abc1}) and (\ref{abc2}),
the only $T$-dependent terms are the $K_{\rm R/I} (E)$ defined above.
Although $K_R(E=0)$ has a power-law $T$ dependence with 
$K_R(E=0)\propto  T^{1/g-1}$, for finite 
energies $E\gg T$, and hence for $T\ll \varepsilon$,
the $T$ dependence is exponentially suppressed since 
the finite level spacing acts as an effective energy 
bias, see Eqs.~(\ref{kernk}) and (\ref{ohmicrate}). 
Similar arguments apply to $K_I(E)$, which can be 
evaluated numerically in a straightforward manner. Although no closed
analytical expression can be given, the overall $T$ dependence follows directly
from these considerations.

\newpage

\begin{thebibliography}{00}
\bibitem{remark4}
There is no need to neglect the dipole-dipole correlations 
$\Lambda_{ij}^{\Sigma}$ in Ref.~\onlinecite{Thorwart02}, 
when the higher-order
contributions are summed up.
\end{thebibliography}
\end{document}